\documentclass[11pt]{article}   	

\usepackage[margin=1in]{geometry}             		
\usepackage[parfill]{parskip} 
\usepackage{graphicx}	
\usepackage{amssymb}
\usepackage{mathtools}
\usepackage{enumerate}
\usepackage{enumitem}
\usepackage{tikz}
\usepackage{amsthm}
\usepackage{bbm}
\usepackage{natbib}
\usepackage{my_style}
\usepackage{hyperref}
\hypersetup{colorlinks,citecolor=blue,urlcolor=blue, linkcolor=blue}
\usepackage{algorithm}
\usepackage{algorithmic}
\usepackage{caption}
\usepackage{subcaption}
\usepackage{authblk}

\title{Detecting Interference in A/B Testing with Increasing Allocation}
\author[1]{Kevin Han}
\author[2]{Shuangning Li}
\author[3]{Jialiang Mao}
\author[1]{Han Wu}
\affil[1]{Department of Statistics, Stanford University, CA, USA}
\affil[2]{Department of Statistics, Harvard University, MA, USA}
\affil[3]{LinkedIn Corporation, Sunnyvale, CA, USA}

\begin{document}
\maketitle

\begin{abstract}
In the past decade, the technology industry has adopted online randomized controlled experiments (a.k.a. A/B testing) to guide product development and make business decisions. In practice, A/B tests are often implemented with increasing treatment allocation: the new treatment is gradually released to an increasing number of units through a sequence of randomized experiments. In scenarios such as experimenting in a social network setting or in a bipartite online marketplace, interference among units may exist, which can harm the validity of simple inference procedures. In this work, we introduce a widely applicable procedure to test for interference in A/B testing with increasing allocation. Our procedure can be implemented on top of an existing A/B testing platform with a separate flow and does not require \textit{a priori} a specific interference mechanism. In particular, we introduce two permutation tests that are valid under different assumptions. Firstly, we introduce a general statistical test for interference requiring no additional assumption. Secondly, we introduce a testing procedure that is valid under a time fixed effect assumption. The testing procedure is of very low computational complexity, it is powerful, and it formalizes a heuristic algorithm implemented already in industry. We demonstrate the performance of the proposed testing procedure through simulations on synthetic data. Finally, we discuss one application at LinkedIn, where a screening step is implemented to detect potential interference in all their marketplace experiments with the proposed methods in the paper.  
\end{abstract}


\section{Introduction}
The technology industry has adopted online randomized controlled experiments, also known as A/B testing, to guide product development and make business decisions \citep{kohavi_deng_2013, kohavi_tang_xu_2020}. In the past decade, firms have developed a dynamic phase release framework in which a new treatment (such as a new product feature) is gradually released to an increasing number of units in the target population through a sequence of randomized experiments \citep{kohavi_tang_xu_2020}. Companies including Google, Microsoft, LinkedIn, and Meta all developed in-house platforms that implement this framework at-scale \citep{tang2010overlapping, kohavi_deng_2013, bakshy2014designing, xu2015infrastructure}. Contrary to the sophisticated engineering design of such platforms, the strategy to analyze A/B testing is relatively simple---often, only the most powerful experiment in the sequence is used to provide a summary of the treatment effect, using tools from classical causal inference assuming independence among test units \citep{imbens_rubin_2015}.

In scenarios such as experimenting in a social network setting or in a bipartite online marketplace, interference among units may exist. Thus, a natural question is whether such interference harms the validity of simple inference procedures. Specific designs have been proposed to test or correct for the interference effects in different applications \citep{saveski2017detecting, EcklesKarrerUgander+2017, ugander_graph_cluster, pouget2019variance, johari2022experimental}. However, these designs are limited to specific applications and often require significant engineering work to implement in parallel to the existing A/B testing infrastructure in most companies. Even when such designs are implemented, their complex nature often results in lower throughput and can slow down the decision process.

In this work, we introduce a widely applicable procedure to test for interference in generic online experiments. The proposed method utilizes data from multiple experiments in the sequence. It can be implemented on top of an existing A/B testing platform with a separate flow and does not require \textit{a priori} the knowledge of the underlying interference mechanism. Once implemented, this test can be run as a standard screening for any A/B test running on the platform. If the test suggests that no interference exists, the experimenter can proceed with classical causal inference analysis with confidence; if the test suggests that some form of interference does exist, the experimenter may need to redesign experiments in a more delicate way. At the platform level, such screening could provide valuable and timely feedback on the choice of designs and help experimenters update development roadmaps accordingly.

\subsection{A motivating example and our contribution}
\label{section:moti_contri}
\begin{figure}
    \centering
    \includegraphics[trim={10pt 0 0 0},clip, width = 0.5\textwidth]{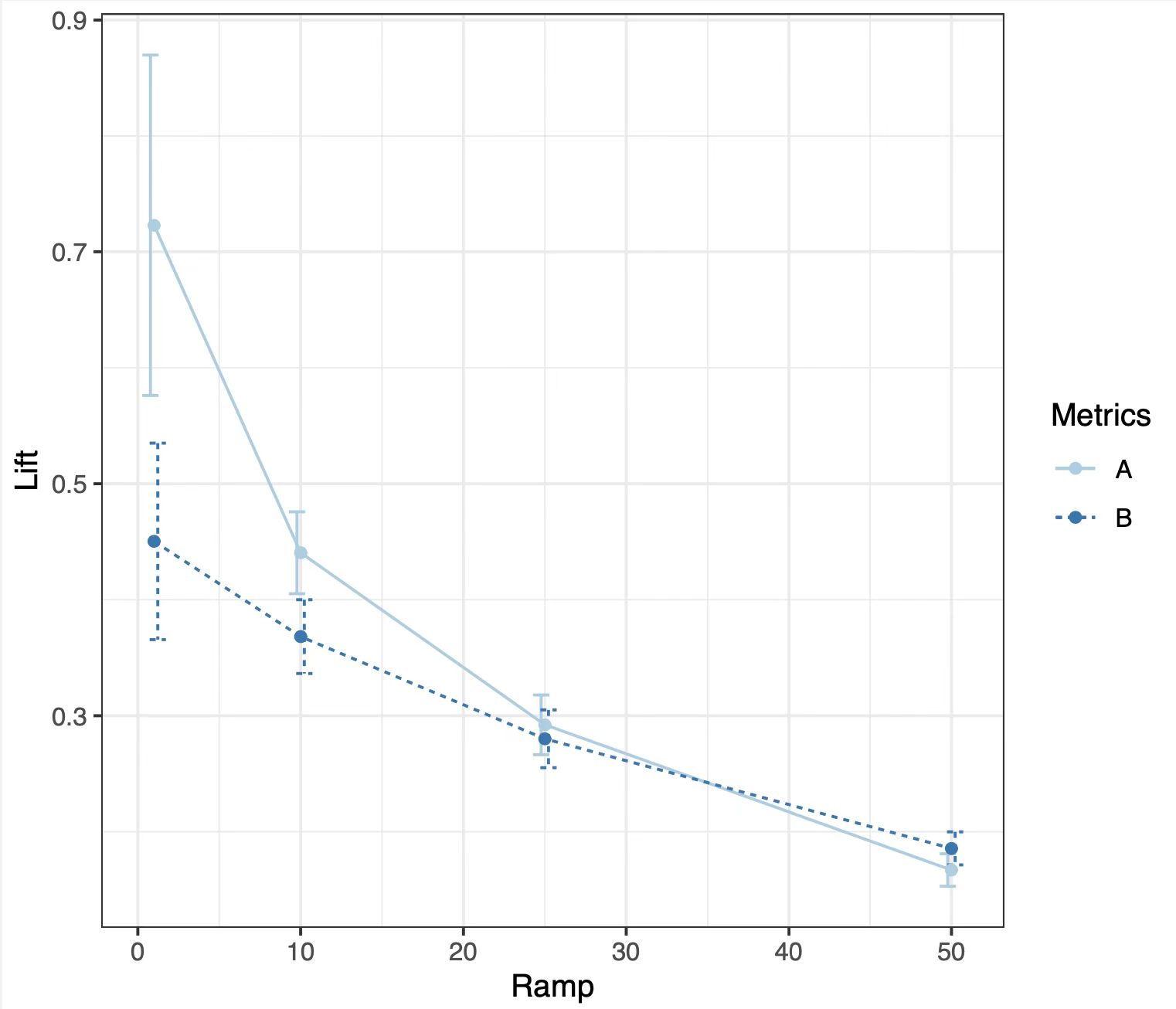}
    \caption{An A/B test implemented by LinkedIn with increasing allocation. On the $x$-axis, we show the percentage of units that are in the treatment group; on the $y$-axis, we show the value of the difference-in-means estimator. Note that A and B stand for different outcome metrics. }
    \label{fig:example_DIM}
\end{figure}

The most straightforward statistical analysis following A/B tests is to compute the difference-in-means estimator, i.e., the difference in the average of outcomes of the treatment group and that of the control group. Under the classical Stable Unit Treatment Value Assumption (SUTVA), which requires that the potential outcomes for any unit do not vary with the treatments assigned to other units, one can easily show that the difference-in-means estimator will be close to the causal effect as long as the sample size is large \citep{imbens_rubin_2015}. This implies that when we compute the difference-in-means estimator for any single randomized experiments in an A/B test with increasing allocation, the value of the estimator should not change by much. However, in some real-world scenarios, we observe drastic change in the difference-in-means estimators throughout the experiments. In Figure \ref{fig:example_DIM}, we show an example from an A/B
test implemented by LinkedIn. In this example, we see that the difference-in-means estimator decreases as the treatment is released to more units. We naturally wonder: What causes this phenomenon? Could it be purely due to randomness? Is the SUTVA assumption violated in this case? 

One plausible explanation for this phenomenon is the existence of interference, i.e., when treatment assigned to one unit may affect observed outcomes for other units. One form of interference is marketplace competition. Imagine a new treatment that can help units perform better in the market. For any particular unit, the treatment brings benefit, but when more of the other units are treated, the other units become more competitive and thus negatively impact the performance of that particular unit. Therefore, in these cases, we often observe that the difference-in-means estimator decreases with treatment probability. Indeed, the experiments in Figure \ref{fig:example_DIM} were run in a setting with marketplace competition. One other common form of interference is through social networks. People's behaviors tend to be positively correlated with those of others connected to them in the network. Think about a treatment that encourages users to comment on a social media platform: users tend to comment more when they see comments from friends. In these cases, we usually observe the difference-in-means estimator increasing with treatment probability.

In practice, however, the structure of interference can be more complicated than the two apparent forms discussed in the above paragraph. Often, experimenters manually examine the difference-in-means plot and decide whether to send the job to other experimentation platforms that deal with interference more carefully. We need a way to formally test whether interference exists. 

In this work, we introduce statistical testing procedures that test for interference in A/B testing with increasing allocation. The methods we propose are scalable and parallelable. They are also agnostic to interference mechanism: even if we have no knowledge of the interference structure, the testing procedure is still valid. Knowledge of the interference structure can, however, be helpful in increasing the power of the testing procedure. We introduce two different testing strategies under different assumptions in Sections \ref{section:general_assu_test} and \ref{section:time_fixed_effect_test}. 
In Section \ref{section:general_assu_test}, we introduce a general statistical test for interference, a test that requires no additional assumptions. The proposed method is inspired by the testing procedure proposed by \citet{athey2018exact}, but it is more powerful than that of \citet{athey2018exact} by making use of multiple experiments. In Section \ref{section:time_fixed_effect_test}, we introduce a testing procedure that is valid under a time fixed effect assumption. The testing procedure is of very low computational complexity, and it is more powerful than the test proposed in Section \ref{section:general_assu_test}. In particular, one special case of this method formalizes a heuristic algorithm discussed above, which decides that interference exists when the difference-in-means estimators are very different.

\subsection{Related work}
\label{section:related_work}
The classical literature on causal inference often assumes that there is no cross-unit interference. When interference presents, many classical inference methods break down. Interest in causal inference with interference started in the social and medical sciences \citep{sobel_2006, hudgens_2008}. Since then, one line of work focuses on estimation and inference of treatment effects under network interference \citep{tchetgen2012, toulis2013estimation, arownow2017, sussman2017elements, basse2018, bhattacharya2020causal, leung2020,  savje2021, savje2021miss, hu2022, li2022}. In order to facilitate estimation, these works either assume that there are special randomization designs or that the interference has some restricted form defined by a given network. Applications to A/B testing are also considered in \cite{ugander_graph_cluster, EcklesKarrerUgander+2017}, and \cite{basse_airoldi_2018}. One assumption implicitly made in these works is that the experiment is conducted only once.
In the multiple experiments regime, \citet{viviano2020experimental} studies the design of two-wave experiments under interference.
\citet{yu2022estimating} and \citet{cortez2022graph} consider estimating the total treatment effects under interference with data from more than two time steps. \cite{bojinov2021} and \cite{Han2021} further investigate the problem in panel experiments. Our work differs from the above works for at least two reasons: (1) instead of focusing on estimation, we focus on testing whether interference exists and (2) we do not need to make additional assumptions in order for the testing procedure to be valid.

In the literature of testing for interference, \cite{bowers_fredrickson_panagopoulos_2013} consider model-based approaches, \citet{pouget2019testing} introduce an experimental design
strategy, and \cite{aronow2012} and \cite{athey2018exact} propose conditional randomization tests restricted to a subset of what they call focal units, and a subset of assignments that make the null hypothesis sharp for focal units. \cite{basse2019} and \cite{puelz2022} further extend this method by using a conditioning mechanism to allow the selection of focal units to depend on the observed treatment assignment. However, none of these works addresses the problem of multiple experiments, and their methods tend to have lower power when directly applied in our setup. To the best of our knowledge, our work is the first to consider testing interference with a sequence of randomized experiments. 

Our work is also related to research on interference in online marketplace experiments (See \citet{basse2016randomization,  fradkin2019simulation, holtz2020reducing, bajari2021multiple, wager2021experimenting, johari2022experimental, li2022interference} among others). This line of work usually requires careful modeling of the market and the interference mechanism. The testing procedure introduced in this paper, in contrast, can be applied to arbitrary forms of interference. 

\section{Problem Setup}
We work in a setting where we run a sequence of A/B tests with increasing allocations.
Formally, suppose that there are $K$ experiments on a population of $n$ units. Let $\pi_k$ be the marginal treatment probability of the $k^{\operatorname{th}}$ experiment. The treatment probabilities satisfy $\pi_1 < \pi_2 < \dots < \pi_K$. 
For each experiment $k \in \cb{1, \dots, K}$ and each unit $i \in \cb{1, \dots, n}$, let
\begin{align*}
    W_{i,k} &:= \textnormal{treatment of unit $i$ assigned in the } k^{\operatorname{th}}\textnormal{ experiment} , \\
    Y_{i,k} &:= \textnormal{outcome of unit $i$ in the $k^{\operatorname{th}}$ experiment}.
\end{align*}
Here we assume that $W_{i,k} \in \cb{0,1}$ is a binary treatment variable and that a value of 1 corresponds to the treatment group while a value of 0 corresponds to the control group. 

The experiments are implemented in the following way. In the first experiment, each unit $i$ is randomly assigned a treatment $W_{i,1}$, where 
\begin{equation}
\label{eqn:generating1}
W_{i,1} \sim \text{Bernoulli}(\pi_1) \text{ independently}.
\end{equation}
In the subsequent experiments, more units are assigned to the treatment group. Specifically, conditioning on the previous treatments, each $W_{i,k}$ is sampled from the following distribution independently:
\begin{equation}
\begin{cases}
\label{eqn:generating2}
&W_{i,k} \sim \text{Bernoulli}\left((\pi_k-\pi_{k-1})/(1-\pi_{k-1})\right), \quad \text{if } W_{i,k-1} = 0;\\
&W_{i,k} = 1, \quad \text{if } W_{i,k-1} = 1.
\end{cases}
\end{equation}
This formulation guarantees that if we look at the $k^{\operatorname{th}}$ experiment alone, then the treatments $W_{i,k}$'s are i.i.d. Bernoulli$(\pi_k)$. 

Let $W_{1:n,1:K}$ be the $n \times K$ treatment matrix and $Y_{1:n,1:K}$ be the $n \times K$ outcome matrix of all units and all experiments. Let $X_i \in \mathbb{R}^d$ be the observed covariates of unit $i$ that do not change over the course of the experiments. Correspondingly, let $X_{1:n} \in \mathbb{R}^{n\times d}$ be the matrix of covariates of all units.

Following the Neyman-Rubin causal model, we assume that potential outcomes $Y_{i,k}(w_{1:n,1:K}) \in \mathbb{R}$ exist for all $w_{1:n,1:K} \in \cb{0, 1}^{n \times K}$ and that the observed outcomes satisfy $Y_{i,k} = Y_{i,k}(W_{1:n,1:K})$.\footnote{In the literature, a \textit{no anticipation effects} assumption is often made in such potential outcome models. The assumption states that the outcome $Y_{i,k}$ depends only on the treatments assigned during and prior to the $k^{\operatorname{th}}$ experiment. With this assumption, the potential outcomes can be written as $Y_{i,k}(w_{1:n,1:k})$ which satisfies $Y_{i,k} = Y_{i,k}(W_{1:n,1:k})$. Here, for simplicity, we keep the original notation.} The goal is to test the following hypothesis:

\begin{hypo}[No cross-unit interference]
\label{hypo:no_cu_interference}
$Y_{i,k} (w_{1:n,1:K}) = Y_{i,k}(\tilde{w}_{1:n,1:K})$ if $w_{i,1:K} = \tilde{w}_{i,1:K}$.
\end{hypo}
The hypothesis states that the outcomes of unit $i$ depend only on the treatments of unit $i$ and not on the treatments of others. We call this hypothesis the no cross-unit interference hypothesis. 

\section{Testing for interference}
In this section, we introduce methods that test for the existence of cross-unit interference. For brevity’s sake, we focus on testing with two experiments. We then discuss further extensions to multiple experiments in Section \ref{section:three_experiments}. 

Naturally, the first question that occurs is how interference might arise. To formalize this, we introduce a notion of \textit{candidate exposure} that captures the potential form of interference. Using domain knowledge, experimenters can specify the candidate exposure, which can vary from application to application.
When we consider user-level data, we have a natural social network. Here experimenters may suspect that a user’s outcome is influenced by treatments of ``friends", i.e., users connected through the social network. And thus in this example, some plausible choices of candidate exposures include the fraction of friends who are treated, and the number of friends who are treated. 
When we consider marketplace competition, advertisers are the subjects of treatment. Here, the sales of an advertiser may be impacted by the treatments of competitors, i.e., advertisers that sell similar products. Hence, in this application, experimenters can choose candidate exposures to be the number of treated advertisers that sell products of the same category, or an average of treatments given to other advertisers weighted by some product similarity metric.
 
Formally, for each experiment $k$ and each unit $i$, we use $H_{i,k} = h_i(W_{-i,k}) \in \mathbb{R}^m$ to denote the candidate exposure. Here $W_{-i,k}$ is the treatments given to all other units except $i$ in the $k^{\operatorname{th}}$ experiment. 
We use the form $h_i(W_{-i,k})$ to emphasize that the candidate exposure depends on other units' treatments. We also write $H_{1:n,k} = (H_{1,k}, H_{2,k}, \dots, H_{n,k})^\top \in \mathbb{R}^{n \times m}$ to reference the candidate exposures of all units.

We want to emphasize that for all the tests introduced below, we do not require the candidate exposure to be correctly specified in order for the tests to be valid. However, the form of the candidate exposure matters for the power of the tests.

We will then move on to test the hypothesis that no interference exists making use of the candidate exposure $H_{i,k}$. In the following sections, we discuss different strategies to test for interference under different assumptions. 

\subsection{Testing under general assumptions}
\label{section:general_assu_test}

We start with a setting where we have access to a dataset from \textit{only one} experiment. Suppose that we collect data on units indexed by $i = 1, \, ..., \, n$, where each unit is randomly assigned to a binary treatment $W_i \in \cb{0, 1}$, 
\begin{equation}
W_i \sim \text{Bernoulli}(\pi) \text{ independently}
\end{equation}
for some $0 \leq \pi \leq 1$.
For each unit, we observe an outcome of interest $Y_i \in \mathbb{R}$ and some covariates $X_i  \in \mathbb{R}^p$. 
\citet{athey2018exact} proposed a method to test for Hypothesis \ref{hypo:no_cu_interference} in this setting.\footnote{The method proposed by \citet{athey2018exact} is more general. Here, we focus on a special case: testing the existence of cross-unit interference in Bernoulli experiments.} We sketch the procedure in Algorithm~\ref{alg:test_one_exp}. 

\begin{algorithm}[htbp]
\caption{\label{alg:test_one_exp} Testing for interference effect (one experiment).}
\begin{flushleft}
{\bf Input:} Dataset $\mathcal{D} = (W_{1:n}, X_{1:n}, Y_{1:n}, H_{1:n})$, exposure function $h$, test statistic $T$. 
\vspace{0.2cm}
\begin{enumerate}
\item Randomly split the data into two folds. Let $\mathcal{I}_{\foc}$ and $\mathcal{I}_{\aux}$ be the index set for the first fold (focal units) and the second fold (auxiliary units). Write the first fold of data as $\mathcal{D}_{\foc} = (W_{\foc}, X_{\foc}, Y_{\foc}, H_{\foc})$ and the second as $\mathcal{D}_{\aux} = (W_{\aux}, X_{\aux}, Y_{\aux}, H_{\aux})$. 
\item Compute a test statistic $T^{(0)} = T(W_{\foc}, X_{\foc}, Y_{\foc}, H_{\foc})$ that captures the importance of $H$ in predicting $Y$. 
\item {\bf For} $b = 1, \dots B$:
\begin{itemize}[label={}]
  
  \item Regenerate treatments for the auxiliary units: $\widetilde{W}^{(b)}_i \sim \operatorname{Bernoulli}(\pi)$ for $i \in \mathcal{I}_{\aux}$.
  \item Recompute the candidate exposure for focal units: $\widetilde{H}_i^{(b)} = h_i( W_{\foc\setminus\cb{i}}, \widetilde{W}_{\aux}^{(b)}))$ for $i \in \mathcal{I}_{\foc}$. 
  \item Recompute the test statistic: $T^{(b)} = T(W_{\foc}, X_{\foc}, Y_{\foc}, \widetilde{H}_{\foc}^{(b)})$. 
\end{itemize}
{\bf End For}
\end{enumerate}

\vspace{0.2cm}
{\bf Output:} The $p$-value 
\begin{equation}
p = \frac{1}{B+1} \p{1 + \sum_{b=1}^B \mathbbm{1}\cb{T^{(0)} \leq T^{(b)}}}. 
\end{equation}
\end{flushleft}
\end{algorithm}

Algorithm~\ref{alg:test_one_exp} requires as input a test statistic $T$ that captures the importance of the candidate exposure $H$ in predicting outcome $Y$. As an illustration, assume for now that $H_i \in \mathbb{R}$. 
One plausible choice of the test statistic $T$ (when $H_i \in \mathbb{R}$) is the following: we run a linear regression of $Y_{\foc} \sim W_{\foc} + X_{\foc} + H_{\foc}$, extract the coefficient of $H_{\foc}$, and take the test statistic $T$ to be the absolute value of the coefficient. We use this regression coefficient statistic as an example to explain the intuition of the algorithm. Under the null hypothesis, the candidate exposure $H$ has no power to predict the outcome $Y$ before or after regenerating treatments, and thus the distribution of the test statistic $T$ will not change after regenerating treatments. Hence, the $\pval$ will be stochastically larger than $\operatorname{Unif}[0,1]$. Under the alternative hypothesis, the behavior of the $\pval$ can be very different. Consider a simple example where $H_i$ is the treatment assigned to the closest friend of unit $i$ and $Y_i = \alpha^\top X_i + \beta W_i + \theta H_i + \epsilon_i$ for some i.i.d. zero mean errors $\epsilon_i$. In this example, the original test statistic $T(W_{\foc}, X_{\foc}, Y_{\foc}, H_{\foc}) \approx \abs{\theta}$ when the sample size is large. However, after regenerating treatments, for each focal unit $i$, if the closest friend of $i$ is among the auxiliary units, then $\widetilde{H}_i$ is marginally a $\operatorname{Bern}(\pi)$ random variable, \textit{independent} of $Y_i$; and hence the distribution of $T(W_{\foc}, X_{\foc}, Y_{\foc}, \widetilde{H}_{\foc}^{(b)})$ will not concentrate around $\abs{\theta}$. In this case, the $\pval$ is far from the $\operatorname{Unif}[0,1]$ distribution. 

In practice, experimenters can use any test statistic $T$ that are suitable for specific applications. For example, if the covariate $X$ is of high dimension, a lasso-type algorithm can be used. One can also run more complicated machine learning algorithms, e.g., random forest and gradient boosting, with $Y$ as a response and $X, W, H$ as predictors, and set the statistic $T$ to be any feature importance statistic of $H$. Just like the choice of candidate exposure $h$, the choice of test statistic $T$ will not hurt the validity of the test, but will largely influence the power of the test. 

\begin{figure}
    \centering
    \includegraphics[width = 0.8\textwidth, trim={0 0 4cm 0},clip]{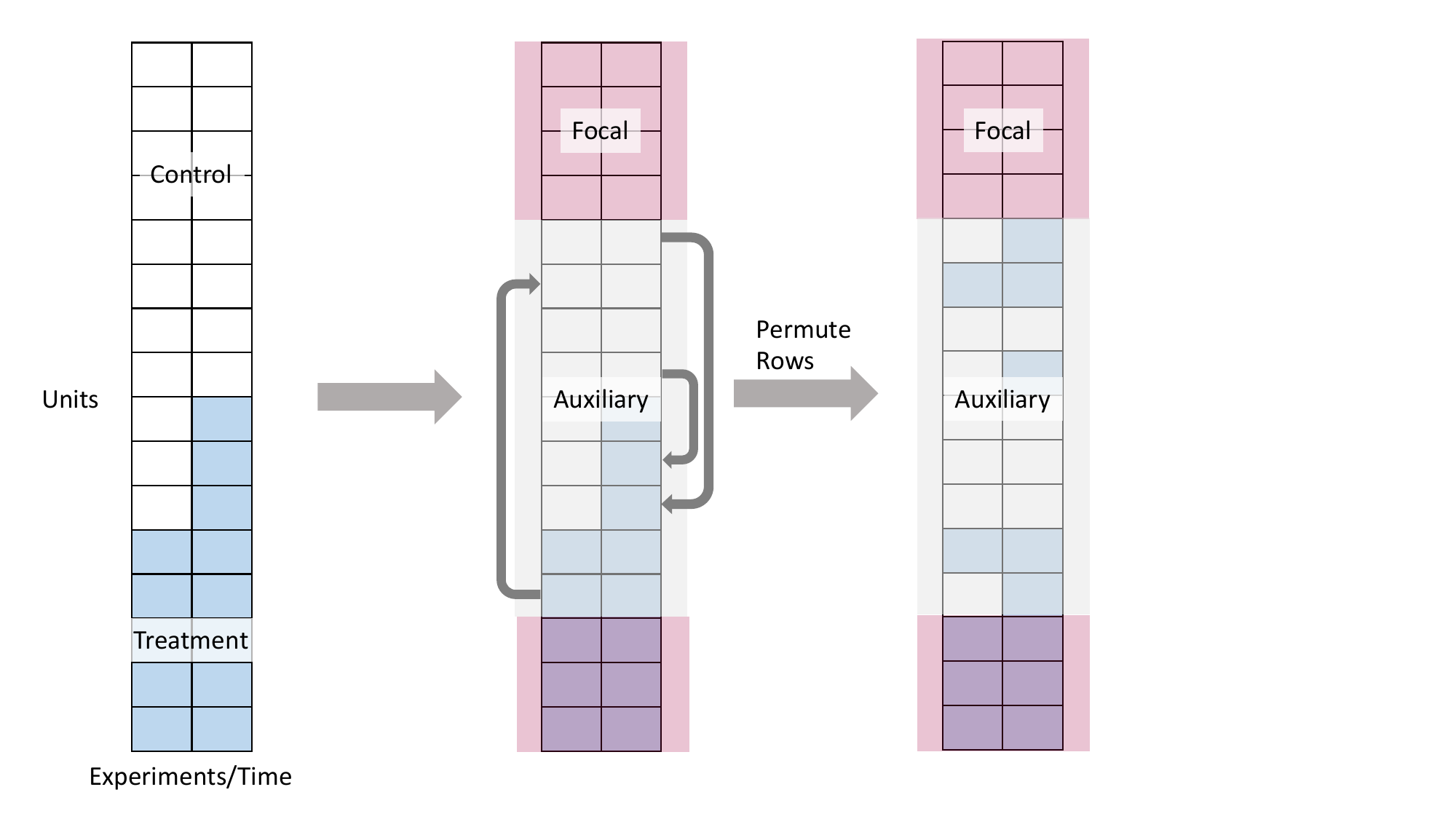}
    \caption{An illustration of Algorithm \ref{alg:test_two_exp}. After selecting the set of focal units and auxiliary units, we randomly permute rows of the treatment matrix and compute test statistics and $p$-values based on the permuted data. }
    \label{fig:illus_vert}
\end{figure}

Then a natural question to ask is whether we can make use of information from multiple experiments to further increase the power of the test. Suppose that we collect data from \textit{two} experiments on the same $n$ units indexed by $i = 1, \dots, n$. In order to increase the power of the previous testing procedure, a natural idea is to reduce the variance in the test statistic computed in Algorithm~\ref{alg:test_one_exp}. To do so, instead of focusing on $Y_{i,2}$ itself, we focus on $Y_{i,2} - Y_{i,1}$. This difference is helpful in removing variance of $Y_i$'s that is shared by $Y_{i,1}$ and $Y_{i,2}$ but cannot be explained by the treatment and covariates. If a unit has some hidden individual characteristics, those characteristics could influence both $Y_{i,1}$ and $Y_{i,2}$ in a similar fashion but may not be well captured by the observed covariates. To make this intuition precise, we present Algorithm~\ref{alg:test_two_exp}, which makes uses of information from two experiments and tests for the existence of interference effect. We have also included an illustration of the algorithm in Figure~\ref{fig:illus_vert}. 

%Before introducing the algorithm, we introduce some more notations. We write $W$ as the matrix with its two columns being $W^1$ and $W^2$. For any index set $\mathcal{I} \in [n]$, we let $W_{\mathcal{I}}$ denote the submatrix of $W$ with row numbers in $\mathcal{I}$. We use similar notations for $H$ and $Y$. 

\begin{algorithm}[t]
\caption{\label{alg:test_two_exp} Testing for interference effect (two experiments).}
\begin{flushleft}
{\bf Input:} Datasets $\mathcal{D}_1 = (W_{1:n,1}, X_{1:n}, Y_{1:n,1}, H_{1:n,1})$, $\mathcal{D}_2 = (W_{1:n,2}, X_{1:n}, Y_{1:n,2}, H_{1:n,2})$, exposure function $h$, test statistic $T$. 
\vspace{0.2cm}
\begin{enumerate}
\item Let $\mathcal{I}_{\operatorname{nc}} = \cb{i: W_{i,1} = W_{i,2}}$ be the set of units whose treatment didn't change over the experiments. Randomly sample a subset of $\mathcal{I}_{\operatorname{nc}}$ of size $n/2$. We call the subset $\mathcal{I}_{\foc}$. Let $\mathcal{I}_{\aux} = [n] \setminus \mathcal{I}_{\foc}$.
\item Take the difference of $Y_{\foc,2}$ and $Y_{\foc,1}$: let $Y^{\operatorname{diff}}_{\foc} = Y_{\foc,2} - Y_{\foc,1}$. Compute a test statistic $T^{(0)} = T(W_{\foc,1:2}, X_{\foc}, Y^{\operatorname{diff}}_{\foc}, H_{\foc,1:2})$ that captures the importance of $H$ in predicting $Y^{\operatorname{diff}}$. 

\item {\bf For} $b = 1, \dots B$:
\begin{itemize}[label={}]
  
  \item Randomly permute treatments for the auxiliary units of the data: $\widetilde{W}^{(b)}_{i,1:2} = W_{\sigma^{(b)}(i),1:2}$ for $i \in \mathcal{I}_{\aux}$, for some permutation $\sigma^{(b)}$ of $\mathcal{I}_{\aux}$.
  
   \item Recompute the candidate exposure for the focal units: $\widetilde{H}_{i,k}^{(b)} = h_i( W_{\foc\setminus\cb{i},k}, \widetilde{W}_{\aux,k}^{(b)})$ for $i \in \mathcal{I}_{\foc}$ and $k \in \cb{1,2}$.

  \item Recompute the test statistic: $T^{(b)} = T(W_{\foc,1:2}, X_{\foc}, Y^{\operatorname{diff}}_{\foc}, \widetilde{H}_{\foc,1:2}^{(b)})$. 
  
  \end{itemize}
{\bf End For}
\end{enumerate}

\vspace{0.2cm}

{\bf Output:} The $p$-value 
\begin{equation}
p = \frac{1}{B+1} \p{1 + \sum_{b=1}^B \mathbbm{1}\cb{T^{(0)}\leq T^{(b)}}}. 
\end{equation}
\end{flushleft}
\end{algorithm} 

Algorithm~\ref{alg:test_two_exp} has a few key differences from Algorithm~\ref{alg:test_one_exp}. First, the choices of focal units are different. In Algorithm~\ref{alg:test_one_exp}, the choice of the focal units cannot depend on the treatment assignments $W_{1:n}$, whereas in Algorithm~\ref{alg:test_two_exp}, the focal units are randomly chosen from those whose treatment didn't change. This specific choice guarantees that the treatment of the $i^{\operatorname{th}}$ unit will not influence the difference of $Y_{i,2}$ and $Y_{i,1}$ much. Second, as mentioned above, in computing the test statistics, $Y^{\operatorname{diff}}$ is used instead of $Y$ itself. As explained above, this helps reduce variance. Third, instead of regenerating treatment, Algorithm~\ref{alg:test_two_exp} permutes the treatment of the auxiliary units. This change is necessary to guarantee the procedure's validity; the choice of focal units depends on the treatment vector, and thus naively regenerating treatments will not give a valid procedure anymore. This will be demonstrated in Section \ref{section:proof_validity}. 

\subsection{Testing with a time fixed effect model}
\label{section:time_fixed_effect_test}
In the previous section, we allow the existence of ``arbitrary time effect". In particular, Hypothesis~\ref{hypo:no_cu_interference} allows the outcome $Y_{i,k}$ to depend on the treatments in other experiments, and does not restrict the relationship among outcomes in different experiments. This brings flexibility and generality, but it could reduce the power of the testing procedures. In this section, we make additional assumptions on the structure of time effect and propose a different testing procedure. 
\begin{assu}[No temporal interference]
\label{assu:no_temp_interf}
$Y_{i,k}(w_{1:n,1:K}) = Y_{i,k}(\tilde{w}_{1:n,1:K})$ if $w_{1:n,k} = \tilde{w}_{1:n,k}$. 
\end{assu}
Assumption \ref{assu:no_temp_interf} states that the outcomes in experiment $k$ depends only on treatments assigned in experiment $k$. In other words, the effect of treatment in one experiment will not carry over to the other experiments. Under Assumption~\ref{assu:no_temp_interf}, we can simplify the notation of potential outcomes: for any $w_{1:n} \in \cb{0,1}^n$, we write $Y_{i,k}(w_{1:n})$ as the potential outcome and assume that the observed outcomes satisfy $Y_{i,k} = Y_{i,k}(W_{1:n,k})$. Note the difference from the previous notation. Previously, we wrote the potential outcomes $Y_{i,k}(w_{1:n, 1:K})$ for any $w_{1:n, 1:K} \in \cb{0,1}^{n \times K}$. Here we focus on the potential outcomes $Y_{i,k}(w_{1:n})$ for any $w_{1:n} \in \cb{0,1}^n$. Following this new notation, we make an additional assumption. 

\begin{assu}[Time fixed effect]
\label{assu:time_fixed_effect}
For any $w_{1:n} \in \cb{0,1}^n$, $i \in \cb{1, \dots, n}$ and $k \in \cb{1, \dots, K}$, $Y_{i,k}(w_{1:n}) = \alpha_i(w_{1:n}) + u_k + \epsilon_{i,k}(w_{1:n})$. The random variables $\epsilon_{i,1}(w_{1:n}), \dots, \epsilon_{i,K}(w_{1:n})$ are zero mean, and are independently and identically distributed, independently of functions $\alpha_{1:n}$,  variables $u_{1:K}$, treatments $W_{1:n, 1:K}$, covariates $X_{1:n}$ and other errors $\epsilon_{j,l}$ for $j \neq i$. 
\end{assu} 

Assumption \ref{assu:time_fixed_effect} assumes a time fixed effect model. The term $u_k$ captures the time effect: some special events may happen when the $k^{\operatorname{th}}$ experiment is implemented, and Assumption \ref{assu:time_fixed_effect} assumes that the effect of such events is shared by all units in the experiments. 
The term $\alpha_i(w)$ captures the individual effect, which could depend on the treatment of unit $i$ as well as treatments of other units. Finally, the terms $\epsilon_{i,k}(w_{1:n})$'s are errors that are i.i.d. across experiments. 

We also note that the commonly used \textit{no temporal effect} assumption is a special case (stronger version) of Assumption \ref{assu:time_fixed_effect}. The no temporal effect assumption assumes that $Y_{i,k}(w_{1:n}) = \alpha_i(w_{1:n}) + \epsilon_{i,k}(w_{1:n})$, where the errors $\epsilon_{i,k}(w_{1:n})$'s are zero mean and i.i.d. across experiments. This corresponds to Assumption \ref{assu:time_fixed_effect} with all time fixed effects $u_k = 0$. The no temporal effect assumption is particularly plausible when all the experiments are implemented within a short period of time, where the distribution of $Y_{i,k}(w_{1:n})$ is not expected to change by much.

Assumption~\ref{assu:no_temp_interf} and Hypothesis~\ref{hypo:no_cu_interference} together state that the outcome $Y_{i,k}$ depend only on the treatment of unit $i$ in experiment $k$. Therefore, under Assumption~\ref{assu:no_temp_interf} and Hypothesis~\ref{hypo:no_cu_interference}, we can further simplify the notation of potential outcomes: for any $w \in \cb{0,1}$ we write $Y_{i,k}(w)$ as the potential outcome and assume that the observed outcomes satisfy $Y_{i,k} = Y_{i,k}(W_{i,k})$.\footnote{Note again the difference with the previous notation. Here we focus on the potential outcomes $Y_{i,k}(w)$ for any $w \in \cb{0,1}$, while we consider $w_{1:n,1:K} \in \cb{0,1}^{n \times K}$ for the most general case and $w_{1:n} \in \cb{0,1}^{n}$ assuming Assumption~\ref{assu:no_temp_interf}.} With this new notation, Assumption~\ref{assu:time_fixed_effect}, together with Assumption~\ref{assu:no_temp_interf} and Hypothesis~\ref{hypo:no_cu_interference}, becomes a new hypothesis: 
\begin{manualhypo}{1'}
For any $w \in \cb{0,1}$, $i \in \cb{1, \dots, n}$ and $k \in \cb{1, \dots, K}$, 
\begin{equation}
\label{eqn:simple_potential_outcome_hypo2}
Y_{i,k}(w) = \alpha_i(w) + u_k + \epsilon_{i,k}(w),
\end{equation}
such that the vectors $\epsilon_{1:n,1}(w), \dots, \epsilon_{1:n,K}(w)$ are independently and identically distributed, independently of functions $\alpha_{1:n}$, vector $u_{1:K}$, treatments $W_{1:n, 1:K}$, covariates $X_{1:n}$ and other errors $\epsilon_{j,l}(w)$ for $l \neq k$.
\end{manualhypo}

This corresponds to the two-way ANOVA in statistics literature \citep{yates1934analysis, fujikoshi1993two} and the two-way fixed effect model in economics literature \citep{bertrand2004much,angrist2009mostly}.

In the previous section, we conduct some permutation tests that permute the data ``vertically", i.e., permute different units. Here with the additional assumptions, we can conduct permutation tests that permute the data ``horizontally", i.e., permute different time points or experiments. 

To motivate the permutation test, consider two units $i$ and $j$. Assume that $i$ has been in the treatment group the whole time, while $j$ has been in the control group the whole time. Under Hypothesis~\hyperref[eqn:simple_potential_outcome_hypo2]{1'}, we have for the first experiment, $Y_{i,1} - Y_{j,1}= \p{\alpha_i(1) + u_1 + \epsilon_{i,1}(1)} - \p{\alpha_j(0) + u_1 + \epsilon_{j,1}(0)} = \alpha_i(1) - \alpha_j(0) +\epsilon_{i,1}(1) -  \epsilon_{j,1}(0)$, and for the second experiment, $Y_{i,2} - Y_{j,2}= \p{\alpha_i(1) + u_2 + \epsilon_{i,2}(1)} - \p{\alpha_j(0) + u_1 + \epsilon_{j,1}(0)} = \alpha_i(1) - \alpha_j(0) +\epsilon_{i,2}(1) -  \epsilon_{j,2}(0)$. Thus,
\begin{equation}
    Y_{i,1} - Y_{j,1} = \alpha_i(1) - \alpha_j(0) +\epsilon_{i,1}(1) -  \epsilon_{j,1}(0) 
    \stackrel{d}{=}
    \alpha_i(1) - \alpha_j(0) +\epsilon_{i,2}(1) -  \epsilon_{j,2}(0) = Y_{i,2} - Y_{j,2}. 
\end{equation}
To put it simply, under Hypothesis \hyperref[eqn:simple_potential_outcome_hypo2]{1'}, $Y_{i,1} - Y_{j,1}$ has the same distribution as $Y_{i,2} - Y_{j,2}$. 
However, when there is cross-unit interference, the two distributions could be different. Consider a simple model: 
\begin{equation}
\label{eqn:motivate_hori_perm}
    Y_{i,k} = W_{i,k} H_{i,k} + \epsilon_{i,k},
\end{equation}
where $H_{i,k}$ is the fraction of neighbors of unit $i$ treated in experiment $k$, and $\epsilon_{i,k}$'s are some i.i.d. zero mean errors. Under this model, $Y_{i,1} - Y_{j,1} = H_{i,1} + \epsilon_{i,1} - \epsilon_{j,1}$ and $Y_{i,2} - Y_{j,2} = H_{i,2} + \epsilon_{i,2} - \epsilon_{j,2}$. When the number of neighbors of unit $i$ is large, by law of large numbers, we have $H_{i,1} \approx \pi_1$ and $H_{i,2} \approx \pi_2$. We can then observe that $Y_{i,1} - Y_{j,1}$ and $Y_{i,2} - Y_{j,2}$ have different distributions; in particular, they have different means.

\begin{algorithm}[t]
\caption{
\label{alg:hori_perm} Testing for interference effect (two experiments, time fixed effect model).}
\begin{flushleft}
{\bf Input:} Datasets $\mathcal{D}_1 = (W_{1:n,1}, X_{1:n}, Y_{1:n,1}, H_{1:n,1})$, $\mathcal{D}_2 = (W_{1:n,2}, X_{1:n}, Y_{1:n,2}, H_{1:n,2})$, matching algorithm $m$, test statistic $T$. 
\vspace{0.2cm}
\begin{enumerate}
\item Let $\mathcal{I}_0 = \cb{i: W_{i,1} = W_{i,2} = 0}$ and $\mathcal{I}_1 = \cb{i: W_{i,1} = W_{i,2} = 1}$. 
\item For each $i$ in $\mathcal{I}_1$, match an index $j \in \mathcal{I}_0$ to $i$ (with no repeat): let $m(i)$ be the matched index of $i$.
Let $\mathcal{I}_m = \cb{m(i): i \in \mathcal{I}_1}$ be the set of matched indices.\footnotemark 
\item 
For each $k \in \cb{1,2}$, compute $Y^{\operatorname{diff}}_{\mathcal{I}_1, k} =  \p{Y_{i,k} - Y_{m(i),k}}_{i \in \mathcal{I}_1}$, which is the vector of differences between the outcomes of the treated units and those of the matched units. \\
Compute a test statistic $T^{(0)} = T(Y^{\operatorname{diff}}_{\mathcal{I}_1,1:2}, X_{\mathcal{I}_m}, H_{\mathcal{I}_m,1:2}, X_{\mathcal{I}_1}, H_{\mathcal{I}_1,1:2})$. 
\item {\bf For} $b = 1, \dots B$:
\begin{itemize}[label={}]
  \item {\bf For} each $i \in \mathcal{I}_1$:
    \begin{itemize}[label={}]
    \item Randomly permute outcomes across experiments: $\widetilde{Y}^{(b)}_{i,k} = Y_{i,\sigma_{i,b}(k)}$ and $\widetilde{Y}^{(b)}_{m(i),k} = Y_{m(i),\sigma_{i,b}(k)}$ for some permutation $\sigma_{i,b}$ of $\cb{1,2}$. 
    \end{itemize}
\textbf{End For}
\item 
Recompute $\widetilde{Y}^{\operatorname{diff}, (b)}_{\mathcal{I}_1,k} =  (\widetilde{Y}^{(b)}_{i,k} - \widetilde{Y}^{(b)}_{m(i),k})_{i \in \mathcal{I}_1}$.\\
Recompute the test statistic: $T^{(b)} = T(\widetilde{Y}^{\operatorname{diff}(b)}_{\mathcal{I}_1,1:2}, X_{\mathcal{I}_m}, 
H_{\mathcal{I}_m,1:2},
X_{\mathcal{I}_1},
H_{\mathcal{I}_1,1:2})$. 
\end{itemize}
\textbf{End For}
\end{enumerate}
\vspace{0.2cm}
{\bf Output:} The $p$-value 
\begin{equation}
p = \frac{1}{B+1} \p{1 + \sum_{b=1}^B \mathbbm{1} \cb{T^{(0)} \leq T^{(b)}}}. 
\end{equation}
\end{flushleft}
\end{algorithm}
\footnotetext{Here we assume that $\abs{\mathcal{I}_1} < \abs{\mathcal{I}_0}$. If $\abs{\mathcal{I}_1} \geq \abs{\mathcal{I}_0}$, we start with $\mathcal{I}_1$ instead. }

\begin{figure}[t]
    \centering
    \includegraphics[width = 0.8\textwidth, trim={0 0 4cm 0},clip]{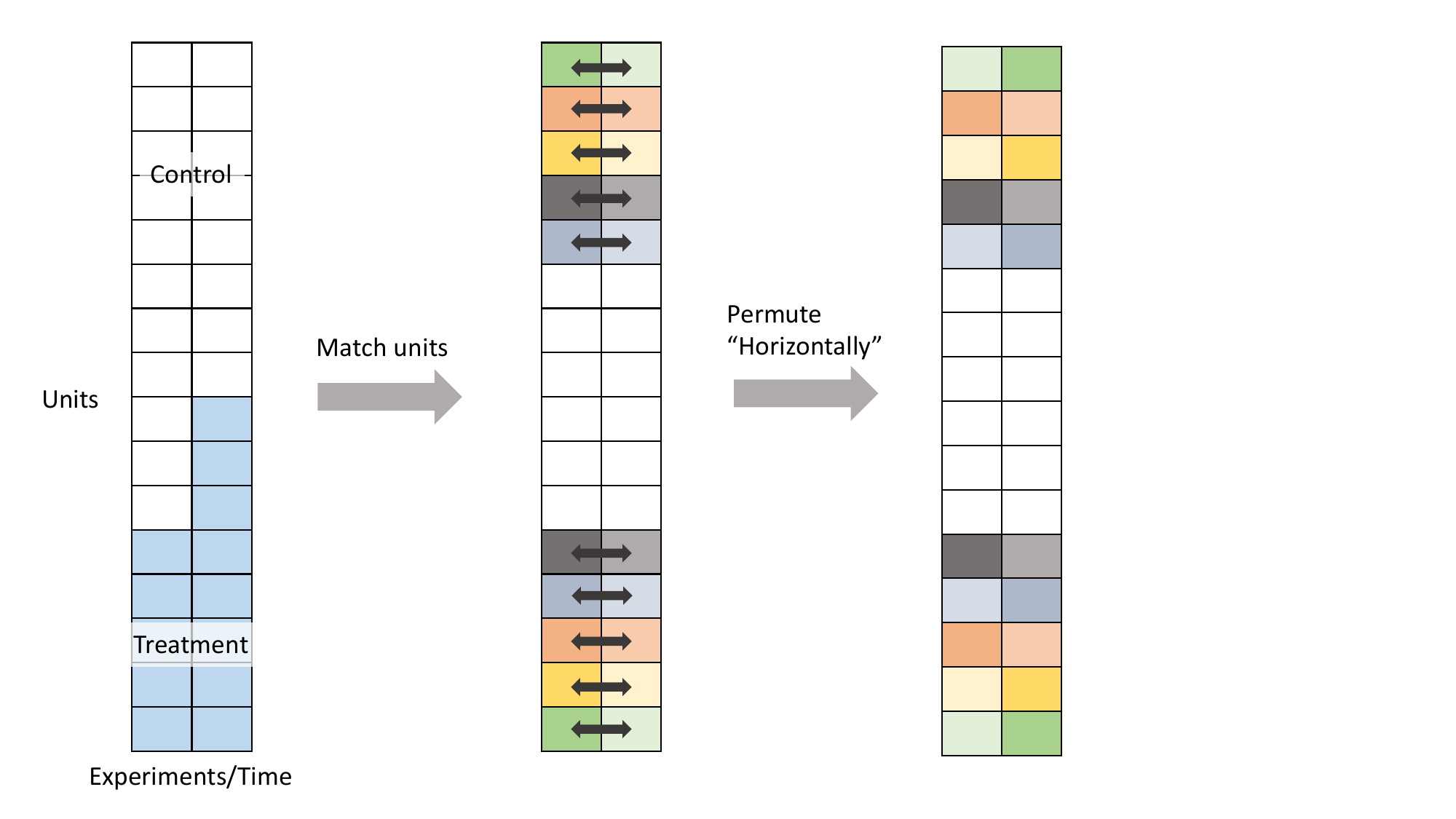}
    \caption{An illustration of Algorithm \ref{alg:hori_perm}. Algorithm \ref{alg:hori_perm} permutes the outcomes horizontally (across experiments), whereas Algorithm \ref{alg:test_two_exp} permutes the treatments vertically (across units).}
    \label{fig:illus_hori}
\end{figure}

Given the above observation, we can conduct a permutation test permuting pairs of $(i,j)$ across experiments. We outline the algorithm in Algorithm \ref{alg:hori_perm}. We also provide an illustration of Algorithm~\ref{alg:hori_perm} in Figure \ref{fig:illus_hori}. 

In Algorithm \ref{alg:hori_perm}, we compare the value of a test statistic to the value of the statistic after permutation. 
One simple choice of test statistic is the difference-in-differences statistic:
\begin{equation}
   T(Y^{\operatorname{diff}}_{\mathcal{I}_1,1:2}, X_{\mathcal{I}_m}, H_{\mathcal{I}_m,1:2}, X_{\mathcal{I}_1}, H_{\mathcal{I}_1,1:2})
= \abs{\operatorname{mean}(Y^{\operatorname{diff}}_{\mathcal{I}_1,2}) -  \operatorname{mean}(Y^{\operatorname{diff}}_{\mathcal{I}_1,1})},
\end{equation}
where $\mathcal{I}_1$ and $\mathcal{I}_m$ are defined in the first step of Algorithm \ref{alg:hori_perm}. 
We use the simple model \eqref{eqn:motivate_hori_perm} discussed above to explain why this choice of statistic is reasonable. Under model \eqref{eqn:motivate_hori_perm}, the difference-in-differences statistic (without absolute value) will be 
\begin{equation}
\operatorname{mean}(Y^{\operatorname{diff}}_{\mathcal{I}_1,2}) -  \operatorname{mean}(Y^{\operatorname{diff}}_{\mathcal{I}_1,1})
\approx \operatorname{mean}(H_{\mathcal{I}_1,2})
- \operatorname{mean}(H_{\mathcal{I}_1,1})
\approx \pi_2 - \pi_1. 
\end{equation}
However, after permutation, the difference-in-differences statistic (without absolute value) will be mean zero. Therefore, $T^{(0)}$ and $T^{(b)}$ will have different distributions and thus the $\pval$ will be far from the  $\operatorname{Unif}[0,1]$ distribution. 

One advantage of this difference-in-differences test statistic is its simplicity. To compute this statistic, there is no need of constructing a candidate exposure or any interference graph, and thus the computation cost of the test statistic is very low. This test statistic is also very intuitive to understand. Recall the motivating example in Section \ref{section:moti_contri}: when the difference-in-means estimators are different, the difference-in-differences test statistic is large. With this test statistic, our algorithm formalizes the intuition of the motivating example in Section \ref{section:moti_contri}. 

The difference-in-differences statistic is not the only one we can choose. Indeed, just as for Algorithms \ref{alg:test_one_exp} and \ref{alg:test_two_exp}, we have full flexibility in choosing the test statistic. For example, we can add covariate adjustment into the test statistics: instead of taking the difference of $\operatorname{mean}(Y^{\operatorname{diff}}_{\mathcal{I}_1,2})$ and $\operatorname{mean}(Y^{\operatorname{diff}}_{\mathcal{I}_1,1})$, we can take the difference of the fitted intercepts after regressing $Y^{\operatorname{diff}}_1$ (and $Y^{\operatorname{diff}}_2$) on $X_{\mathcal{I}_m}$ and $X_{\mathcal{I}_1}$. We can also bring the candidate exposure $H$ into the picture. For example, we can similarly define $H^{\operatorname{diff}}_{\mathcal{I}_1,k} =  \p{H_{i,k} - H_{m(i),k}}_{i \in \mathcal{I}_1}$ for $k \in \cb{1,2}$. Then one plausible test statistic (when $H_{i,k} \in \mathbb{R}$) is the following:
\begin{equation}
   T(Y^{\operatorname{diff}}_{\mathcal{I}_1,1:2}, X_{\mathcal{I}_m}, H_{\mathcal{I}_m,1:2}, X_{\mathcal{I}_1}, H_{\mathcal{I}_1,1:2})
= \abs{\Corr{Y^{\operatorname{diff}}_{\mathcal{I}_1,2}-  Y^{\operatorname{diff}}_{\mathcal{I}_1,1},
H^{\operatorname{diff}}_{\mathcal{I}_1,2}-  H^{\operatorname{diff}}_{\mathcal{I}_1,1}
}}. 
\end{equation}
%We will discuss in more detail the advantages and disadvantages of different choices of statistics from a practical perspective in Section \ref{section:simulation}. 

Finally, we want to comment on the matching algorithm $m$ used in Algorithm \ref{alg:hori_perm}. We would first like to stress that as long as the matching algorithm only looks at the covariates $X$, the test will be valid regardless of the quality of matching. In the most extreme case, we can simply conduct a random matching, and the test will remain valid. More ideally, we would hope each $i$ is matched to an $m(i)$ such that $X_i$ is close to $X_{m(i)}$. This matching step helps reduce variance due to the covariates and thus increase the power of the test. In the causal inference literature, matching algorithms have been widely studied \citep{rubin1973matching, stuart2010matching}, and we recommend that experimenters choose from existing algorithms based on their needs and the computational resources available.

\subsection{Usage of graphs of experimental units}
In implementing the previously proposed algorithms, we often find it helpful to construct a graph of the $n$ experimental units. Formally, let $G = (V,E)$, with vertex set $V = \cb{1,2, \dots, n}$ and edge set $E = \cb{E_{ij}}_{i,j = 1}^n$. 
We will discuss a few different ways of using graphs to test and learn interference structure. 

\paragraph{Interference graph.}
A graph can be constructed to model interference and to help compute candidate exposure. We call such a graph an \textit{interference graph}. 
When experimental units are users, it is plausible to assume that a user's behavior is mostly influenced by friends in a social network. In this case, we can simply take the interference graph to be the social network, i.e., we set $E_{ij} = 1$ if user $i$ and $j$ are friends on the social network. With this graph, many candidate exposures can be computed easily: number of treated friends $H_{i,k}^{\operatorname{numFrds}} = \sum_{j: E_{ij} = 1} W_{j,k}$, fraction of friends that are treated $H_{i,k}^{\operatorname{fracFrds}} = \sum_{j: E_{ij} = 1} W_{j,k}/\abs{\cb{j: E_{ij} = 1}}$, number of treated two-hop friends $H_{i,k}^{\operatorname{num2Frds}} = \sum_{l: \exists j \textnormal{ s.t.} E_{ij}E_{jl} = 1} W_{j,k}$. 
The interference graph can be constructed differently in other settings. When experimental units are advertisers, there is no natural social network. However, we can construct a ``competition network" based on the similarity of the covariates. For a similarity measure $s$ and a threshold $\epsilon$, we can define $E_{ij} = \one\cb{s(X_i, X_i) \geq \epsilon}$. Such a graph reflects that an advertiser is mostly influenced by its competitors, especially those that are similar to it. Candidate exposures can then be computed based on this interference graph: number of treated competitors $H_{i,k}^{\operatorname{numCpt}} = \sum_{j: E_{ij} = 1} W_{j,k}$, weighted average of competitors' treatments: $H_{i,k}^{\operatorname{wAvgCpt}} = \sum_{j: E_{ij} = 1} s(X_i, X_i) W_{j,k}$. 

The interference graph also helps experimenters to understand the nature of interference. Imagine we have two different interference graphs $G_1$ and $G_2$ and we apply the testing procedure separately using $G_1$ and $G_2$. If we observe a much smaller $\pval$ for the procedure using $G_1$ than that we obtain using $G_2$, then we have some evidence suggesting that the interference in the form of $G_1$ is much stronger than in that of $G_2$. In particular, the units that are connected to unit $i$ in $G_1$ might be the most influential in impacting the outcome of unit $i$. This kind of analysis, though not fully rigorous, can help experimenters to build better intuitions for modelling in subsequent analysis. For example, once the interference effect is statistically significant, experimenters may consider re-running experiments with a cluster randomized controlled trial. Understanding the structure of interference can be helpful in constructing better clusters. 

\paragraph{Graph in matching.}
A graph can also be helpful in the matching step in Algorithm \ref{alg:hori_perm}. In the causal inference literature, matched pairs are often constructed using a minimum cost flow algorithm on a bipartite graph with treated units on one side and control units on the other side \citep{rosenbaum1989optimal, hansen2006optimal}. Here, the cost of flow from unit $i$ to $j$ can be defined as some dissimilarity metric between $X_i$ and $X_j$. For example, the Mahalanobis distance is a common choice of such a dissimilarity metric \citep{rubin1980bias}. The bipartite graph may not always be a complete bipartite graph: sometimes a caliper can be applied to the graph resulting in the removal of edges. A caliper based on covariates limits with which a unit can be paired \citep{mahmood2018performance}.\footnote{In the observational study literature, calipers are often applied on the propensity score \citep{cochran1973controlling, rosenbaum1985constructing}. Here we are in an experimental setting instead, where the propensity score is known, and it is the same for all units.} 
For example, researchers may only want advertisers to be matched/paired with advertisers who sell products of the same category; in such cases, there is an edge between $i$ and $j$ only if they sell products of the same category. 

Interestingly, calipered graphs may correspond to the interference graph introduced in the above section, and thus we only need to construct the graph once and use it in both the step of computing candidate exposure and the step of matching. This is especially relevant in a market competition application: a company is expected to be mostly influenced by companies selling similar products, and thus we put edges in the interference graph; in the meantime, we would like to match companies selling similar products, and thus we put edges in the bipartite graph used in matching.

\subsection{Aggregating \texorpdfstring{$p$}{p}-values}
One issue with the algorithms above proposed is that randomly splitting the data (Algorithms~\ref{alg:test_one_exp} and \ref{alg:test_two_exp}) or the random matching step (Algorithm \ref{alg:hori_perm}) can inject randomness into the $\pval$. In order to derandomize the procedure, we can run the algorithms many times and aggregate the $\pvals$. Since the $\pvals$ can be arbitrarily dependent on each other, we cannot use Fisher's method to aggregate the $\pvals$, which requires independence \citep{fisher1925statistical}. Some possible ways include, e.g., setting $p = 2 \sum p_i /n$ (See \citet{vovk2020combining} for more details). 

In the previous section, we discuss the usage of an interference graph in constructing candidate exposure. In practice, experimenters may construct several interference graphs with different sparsity or structure. We can make use of information from different graphs and construct an ``aggregated $\pval$". We can run the algorithms separately for each graph, and compute an ``aggregated test statistic". For example, we can choose $T^{\operatorname{aggre}} = \sum_m T(G_m)$, where $G_m$ is the $m^{\operatorname{th}}$ interference graph considered. Then we can compute an aggregated $\pval$ in the following way:
\begin{equation}
p^{\operatorname{aggre}} = \frac{1}{B+1} \p{1 + \sum_{m=1}^B \mathbbm{1}\cb{T^{\operatorname{aggre}} \leq T^{\operatorname{aggre} (b)}}}. 
\end{equation}

\subsection{Extension to three or more experiments}
\label{section:three_experiments}
More generally, experiments may be conducted more than two times. Formally, suppose that we run $K$ experiments where treatments are randomly assigned according to \eqref{eqn:generating1} and \eqref{eqn:generating2}. 
To test for interference, we can adopt a similar strategy as in Section \ref{section:general_assu_test}. We outline the general algorithm in Algorithm~\ref{alg:test_three_exp}. We note that Algorithm~\ref{alg:test_two_exp} is a special case of Algorithm~\ref{alg:test_three_exp}. In practice, we recommend computing the test statistic using the difference of outcomes between experiments (as emphasized in Algorithm~\ref{alg:test_two_exp}), since this helps remove common variance shared by outcomes in the experiments. One example of such statistic is the following. 
\begin{equation}
\label{eqn:test_stat_corr}
    T(W_{\foc, 1:K},X_{\foc},Y_{\foc, 1:K},H_{\foc, 1:K}) = \sum_{(k,l): k \neq l}\abs{\Corr{Y_{\foc, k} - Y_{\foc, l}, H_{\foc, l} - H_{\foc, l}}}.
\end{equation}

\begin{algorithm}
\caption{\label{alg:test_three_exp} Testing for interference effect (multiple experiments).}
\begin{flushleft}
{\bf Input:} Datasets $\mathcal{D}_k = (W_{1:n,k}, X_{1:n}, Y_{1:n,k}, H_{1:n,k})$ for $k = 1, \dots, K$, exposure function $h$, test statistic $T$. 
\vspace{0.2cm}
\begin{enumerate}
\item Let $\mathcal{I}_{\operatorname{nc}} = \cb{i: W_{i,1} = \dots = W_{i,K}}$ be the set of units whose treatment didn't change over the experiments. Randomly sample a subset of $\mathcal{I}_{\operatorname{nc}}$ of size $n/2$. We call the subset $\mathcal{I}_{\foc}$. Let $\mathcal{I}_{\aux} = [n] \setminus \mathcal{I}_{\foc}$.
\item Compute a test statistic $T^{(0)} = T(W_{\foc,1:K}, X_{\foc}, Y_{\foc,1:K}, H_{\foc,1:K})$ that captures the importance of $H$ in predicting $Y$.
\item {\bf For} $b = 1, \dots B$:
\begin{itemize}[label={}]
  
  \item Randomly permute treatments for the auxiliary units of the data: $\widetilde{W}^{(b)}_{i,1:K} = W_{\sigma^{(b)}(i),1:K}$ for $i \in \mathcal{I}_{\aux}$, for some permutation $\sigma^{(b)}$ of $\mathcal{I}_{\aux}$.  
   \item Recompute the candidate exposure for the focal units: $\widetilde{H}_{i,k}^{ (b)} = h_i(W_{\foc\setminus\cb{i},k}, \widetilde{W}_{\aux,k}^{(b)})$, for $i \in \mathcal{I}_{\foc}$ and $k \in \cb{1,2, \dots, K}$. 
   
  \item Recompute the test statistic: $T^{(b)} = T(W_{\foc,1:K}, X_{\foc}, Y_{\foc,1:K}, \widetilde{H}_{\foc,1:K}^{(b)})$. 
\end{itemize}
{\bf End For}
\end{enumerate}

\vspace{0.2cm}

{\bf Output:} The $p$-value 
\begin{equation}
p = \frac{1}{B+1} \p{1 + \sum_{b=1}^B \mathbbm{1}\cb{T^{(0)}\leq T^{(b)}}}. 
\end{equation}
\end{flushleft}
\end{algorithm}

\begin{figure}
    \centering
    \includegraphics[width = 0.8\textwidth, trim={0 0 4cm 0},clip]{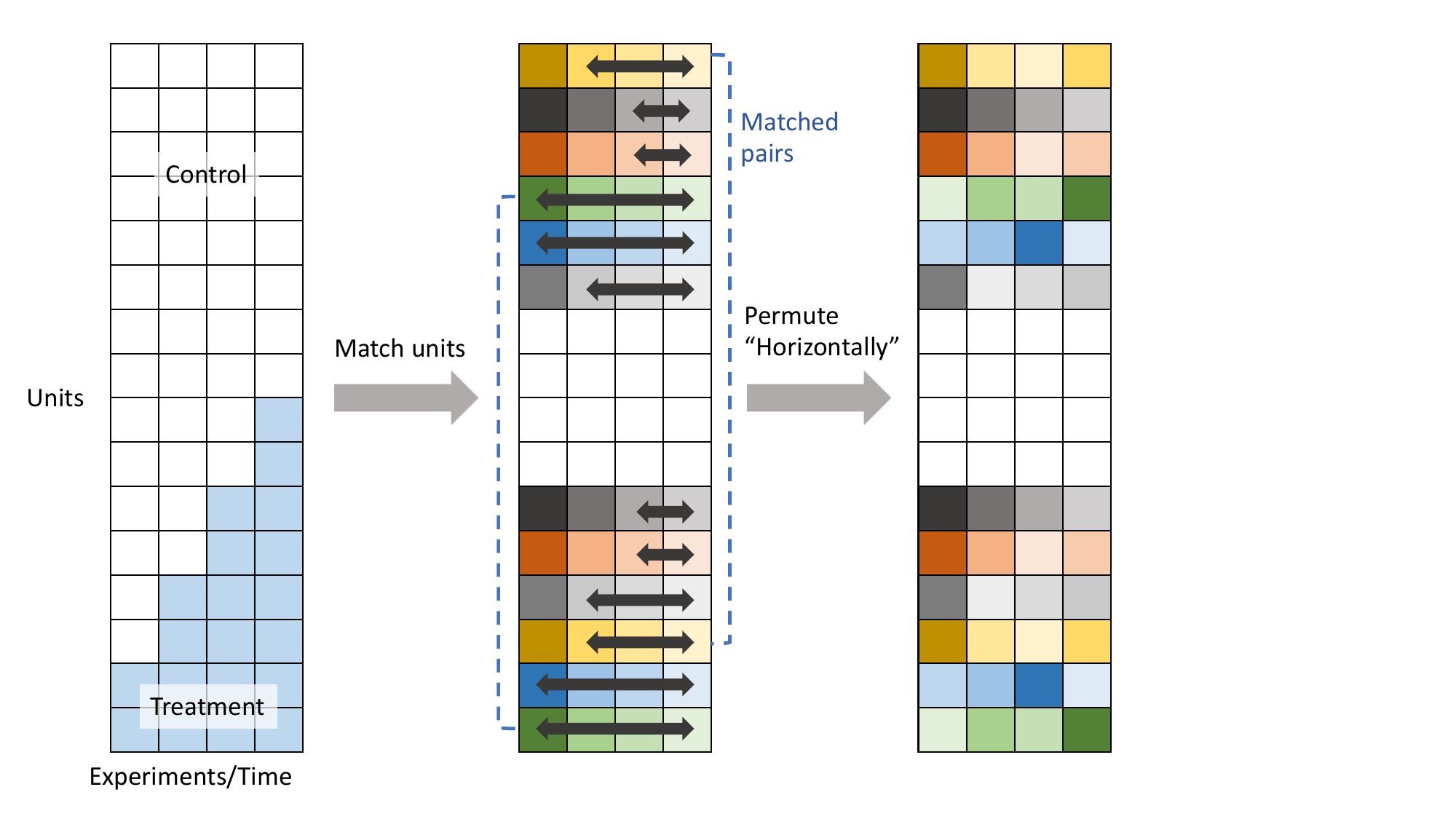}
    \caption{An illustration of Algorithm \ref{alg:hori_perm_multi}. Pairs of units are matched and the outcomes of paired units are permuted together across experiments. Test statistics and the $p$-value are then obtained based on the permuted data. } 
    \label{fig:hori_perm_multi}
\end{figure}

If we assume a time fixed effect model as in Section \ref{section:time_fixed_effect_test}, we can then extend Algorithm \ref{alg:hori_perm} to settings with more experiments. We outline the algorithm in Algorithm \ref{alg:hori_perm_multi}. Again, we note that Algorithm \ref{alg:hori_perm} is a special case of Algorithm \ref{alg:hori_perm_multi}. Algorithm \ref{alg:hori_perm_multi} allows permutation over more experiments than Algorithm \ref{alg:hori_perm} does. In particular, if unit $i$ is treated in experiments $K_1, K_1 + 1, \dots, K$, then the algorithm permutes outcome for unit $i$ and its matched unit over experiments $K_1, K_1 + 1, \dots, K$. Permuting over more experiments helps the test to leverage information from more experiments and thus increases power of the test. We have included an illustration of this algorithm in Figure~\ref{fig:hori_perm_multi}.

\begin{algorithm}
\caption{
\label{alg:hori_perm_multi} Testing for interference effect (multiple experiments, time fixed effect model).}
\begin{flushleft}
{\bf Input:} Datasets $\mathcal{D}_k = (W_{1:n,k}, X_{1:n}, Y_{1:n,k}, H_{1:n,k})$ for $k = 1, \dots, K$, matching algorithm $m$, test statistic $T$. 
\vspace{0.2cm}
\begin{enumerate}
\item Let $\mathcal{I}_0 = \cb{i: W_{i,1} = \dots = W_{i,k} = 0}$ be the set of units that are in the control group in all experiments. Let $\mathcal{I}_1 = \cb{i: W_{i,K-1} = W_{i,K} = 1}$ be the set of units that are in the treatment group in the last two experiments (i.e. units that are treated in at least two experiments).
\item For each $i$ in $\mathcal{I}_1$, match an index $j \in \mathcal{I}_0$ to $i$ (with no repeat): let $m(i)$ be the matched index of $i$.
Let $\mathcal{I}_m = \cb{m(i): i \in \mathcal{I}_1}$ be the set of matched indices.\footnotemark 
\item 
For each $k \in \cb{1, \dots, K}$, compute $Y^{\operatorname{diff}}_{\mathcal{I}_1,k} =  \p{Y_{i,k} - Y_{m(i),k}}_{i \in \mathcal{I}_1}$, which is the vector of differences between the outcomes of the units in $\mathcal{I}_0$ and those of the matched units. \\
Compute a test statistic $T^{(0)} = T(Y^{\operatorname{diff}}_{\mathcal{I}_1,1:K}, X_{\mathcal{I}_m}, H_{\mathcal{I}_m,1:K}, X_{\mathcal{I}_1}, H_{\mathcal{I}_1,1:K})$. 
\item {\bf For} $b = 1, \dots B$:
\begin{itemize}[label={}]
  \item {\bf For} each $i \in \mathcal{I}_1$:
    \begin{itemize}[label={}]
    \item Let $S_i = \cb{k: W_{i,k} = 1}$ be the set of experiments in which unit $i$ is treated. 
    \item Randomly permute outcomes across $S_i$: $\widetilde{Y}^{(b)}_{i,k} = Y_{i,\sigma_{i,b}(k)}$ and $\widetilde{Y}^{(b)}_{m(i),k} = Y_{m(i),\sigma_{i,b}(k)}$ for all $k \in S_i$, where $\sigma_{i,b}$ is a random permutation of $S_i$. 
    \end{itemize}
\textbf{End For}
\item 
Recompute $\widetilde{Y}^{\operatorname{diff}, (b)}_{\mathcal{I}_1,k} =  (\widetilde{Y}^{(b)}_{i,k} - \widetilde{Y}^{(b)}_{m(i),k})_{i \in \mathcal{I}_1}$.\\
Recompute the test statistic: $T^{(b)} = T(\widetilde{Y}^{\operatorname{diff}(b)}_{\mathcal{I}_1,1:K}, X_{\mathcal{I}_m}, 
H_{\mathcal{I}_m,1:K},
X_{\mathcal{I}_1},
H_{\mathcal{I}_1,1:K})$. 
\end{itemize}
\textbf{End For}
\end{enumerate}
\vspace{0.2cm}
{\bf Output:} The $p$-value 
\begin{equation}
p = \frac{1}{B+1} \p{1 + \sum_{b=1}^B \mathbbm{1} \cb{T^{(0)} \leq T^{(b)}}}. 
\end{equation}
\end{flushleft}
\end{algorithm}
\footnotetext{Here we assume that $\abs{\mathcal{I}_0} \geq n/2$.}

\section{Validity of the testing procedures}
\label{section:proof_validity}
In this section, we establish validity of the above proposed algorithms. We make use of the following theorem in \citet[Theorem 2]{hemerik2018exact, hemerik2018false}. 

\begin{theo}[Random permutations]
\label{theo:key_perm}
Let $A_1, A_2, \dots, A_n \in \mathcal{A}$ be $n$ random variables. Let $\mathcal{S}_n$ denote the set of all permutations on $[n]$. Assume that
\begin{enumerate}
    \item $G \subset \mathcal{S}_n$ is a subgroup;
    \item For any $\sigma \in G$, $A = (A_1, \dots, A_n) \stackrel{d}{=} (A_{\sigma(1)}, \dots, A_{\sigma(n)}) = A_{\sigma}$. 
\end{enumerate}
If $\sigma_1, \dots, \sigma_B$ are drawn independently uniformly from $G$, then for any test statistic $T$, the $\pval$ 
\begin{equation}
p = \frac{1}{B+1} \p{1 + \sum_{b=1}^B \mathbbm{1}\cb{T(A)\leq T(A_{\sigma})}}
\end{equation}
satisfies
\begin{equation}
    \PP{p \leq \alpha} \leq \alpha. 
\end{equation}
for any $\alpha \in (0,1)$. 
\end{theo}

We start with establishing the validity of Algorithms \ref{alg:test_one_exp}, \ref{alg:test_two_exp} and \ref{alg:test_three_exp} under general assumptions. 

\begin{theo}
Assume that the treatments are assigned according to rules defined in \eqref{eqn:generating1} and \eqref{eqn:generating2}. Under Hypothesis \ref{hypo:no_cu_interference}, the $p$-values produced by Algorithms \ref{alg:test_one_exp}, \ref{alg:test_two_exp} and \ref{alg:test_three_exp} are valid in the following sense: for any $\alpha \in (0,1)$,
\begin{equation}
\label{eqn:theorem1}
    \PP{p \leq \alpha} \leq \alpha. 
\end{equation}
\end{theo}

\begin{proof}
Algorithm \ref{alg:test_one_exp} has been shown to provide valid $\pvals$ in \citet{athey2018exact}.  
Since Algorithm~\ref{alg:test_two_exp} is a special case of Algorithm~\ref{alg:test_three_exp}, it suffices to prove that the $\pvals$ produced by Algorithm~\ref{alg:test_three_exp} are valid. 
We will be making use of Theorem \ref{theo:key_perm} to show the result. 

We start by noting that since $H_{\foc,1:K}$ is a function of $W_{\foc,1:K}$ and $W_{\aux,1:K}$, the test statistic $T(W_{\foc,1:K}, X_{\foc}, Y_{\foc,1:K}, H_{\foc,1:K})$ can be rewritten as 
\begin{equation}
T(W_{\foc,1:K}, X_{\foc}, Y_{\foc,1:K}, H_{\foc,1:K})
= \check{T}(W_{\foc,1:K}, X_{\foc}, Y_{\foc,1:K}, W_{\aux,1:K})
\end{equation}
for some function $\check{T}$. Thus, we can also rewrite 
\begin{equation}
T(W_{\foc,1:K}, X_{\foc}, Y_{\foc,1:K}, \widetilde{H}_{\foc,1:K}^{(b)})
= \check{T}(W_{\foc,1:K}, X_{\foc}, Y_{\foc,1:K}, \widetilde{W}_{\aux,1:K}^{(b)}).
\end{equation} By construction, $\widetilde{W}_{\aux,1:K}^{(b)}$ is a random permutation of the rows of $W_{\aux,1:K}$. Thus, we can take the permutation group $G$ to be the set of all permutation on $\mathcal{I}_{\aux}$. By Theorem \ref{theo:key_perm}, it suffices to establish that
\begin{equation}
W_{\sigma(\aux),1:K} \mid W_{\foc,1:K}, X_{\foc}, Y_{\foc,1:K}, \mathcal{I}_{\foc}
\stackrel{d}{=}
W_{\aux,1:K} \mid W_{\foc,1:K}, X_{\foc}, Y_{\foc,1:K}, \mathcal{I}_{\foc},
\end{equation}
for any permutation $\sigma(\aux)$ on $\mathcal{I}_{\aux}$. The above is equivalent to 
\begin{equation}
\label{eqn:joint_same}
\begin{split}
&\p{W_{\sigma(\aux),1:K}, W_{\foc,1:K}, X_{\foc}, Y_{\foc,1:K}, \one\cb{\mathcal{I}_{\foc} = \mathcal{I}_{\fix}}}\\
&\qquad \qquad \stackrel{d}{=}
\p{W_{\aux,1:K}, W_{\foc,1:K}, X_{\foc}, Y_{\foc,1:K}, \one\cb{\mathcal{I}_{\foc} = \mathcal{I}_{\fix}}},
\end{split}
\end{equation}
for any fixed subset $\mathcal{I}_{\fix} \subset [n]$ of size $n/2$. Let $\mathcal{I}_{\fixc} = [n] \setminus \mathcal{I}_{\foc}$. Then, under the null hypothesis \ref{hypo:no_cu_interference}, 
\begin{equation}
\begin{split}
&p\p{W_{\aux,1:K}, W_{\foc,1:K}, X_{\foc}, Y_{\foc,1:K}, \one\cb{\mathcal{I}_{\foc} = \mathcal{I}_{\fix}}}\\
&\qquad \qquad = p\p{W_{\fixc,1:K}, W_{\fix,1:K}, X_{\fix}, Y_{\fix,1:K}, \one\cb{\mathcal{I}_{\foc} = \mathcal{I}_{\fix}}}\\
&\qquad \qquad = p(W_{\fixc,1:K})p(W_{\fix,1:K}, X_{\fix}, Y_{\fix,1:K})\PP{\mathcal{I}_{\foc} = \mathcal{I}_{\fix} \mid W_{\fixc,1:K},W_{\fix,1:K} },
\end{split}
\end{equation}
where the last line follows from the no cross-unit interference hypothesis and the fact that treatments are sampled independently across units. Note also that permuting $\mathcal{I}_{\fixc}$ will not change the selection probability of the focal units, i.e., 
$\PP{\mathcal{I}_{\foc} = \mathcal{I}_{\fix} \mid W_{\fixc},W_{\fix} } = \PP{\mathcal{I}_{\foc} = \mathcal{I}_{\fix} \mid W_{\sigma(\fixc)}, W_{\fix}}$, and thus
\begin{equation}
\begin{split}
&p(W_{\fixc,1:K})p(W_{\fix,1:K}, X_{\fix}, Y_{\fix,1:K})\PP{\mathcal{I}_{\foc} = \mathcal{I}_{\fix} \mid W_{\fixc,1:K},W_{\fix,1:K} }\\
&\qquad \qquad = p(W_{\sigma(\fixc),1:K})p(W_{\fix,1:K}, X_{\fix}, Y_{\fix,1:K})\PP{\mathcal{I}_{\foc} = \mathcal{I}_{\fix} \mid W_{\sigma(\fixc),1:K},W_{\fix,1:K} }\\
&\qquad \qquad = p\p{W_{\sigma(\fixc),1:K}, W_{\fix,1:K}, X_{\fix}, Y_{\fix,1:K}, \one\cb{\mathcal{I}_{\foc} = \mathcal{I}_{\fix}}}\\
&\qquad \qquad = p\p{W_{\sigma(\aux),1:K}, W_{\foc,1:K}, X_{\foc}, Y_{\foc,1:K}, \one\cb{\mathcal{I}_{\foc} = \mathcal{I}_{\fix}}},
\end{split}
\end{equation}
and thus proving \eqref{eqn:joint_same}. 
\end{proof}

%%%%%%%%%%%%%%%%%%%%%%%%%%%%%%%%%

\begin{theo}[Time fixed effect model]
Assume that the treatments are assigned according to rules defined in \eqref{eqn:generating1} and \eqref{eqn:generating2}. Under Assumptions \ref{assu:no_temp_interf}- \ref{assu:time_fixed_effect} and Hypothesis \ref{hypo:no_cu_interference}, the $p$-values produced by Algorithms \ref{alg:hori_perm} and \ref{alg:hori_perm_multi} are valid in the following sense: for any $\alpha \in (0,1)$,
\begin{equation}
    \PP{p \leq \alpha} \leq \alpha. 
\end{equation}
\end{theo}

\begin{proof}
Algorithm \ref{alg:hori_perm} is a special case of Algorithm \ref{alg:hori_perm_multi}, and thus we will only work with Algorithm \ref{alg:hori_perm_multi} here. 
We will again make use of Theorem \ref{theo:key_perm} to show the result. 

By construction, the elements in $\widetilde{Y}^{\operatorname{diff},(b)}_{\mathcal{I}_1,1:K}$ are a random permutation of the elements in $Y^{\operatorname{diff}}_{\mathcal{I}_1,1:K}$.
The allowed permutations in Algorithm \ref{alg:hori_perm_multi} clearly form a group. 
Specifically, the allowed permutations are defined by $\sigma = (\sigma_i)_{i \in \mathcal{I}_1}$, where each $\sigma_i$ is a permutation of $S_i = \cb{k : W_{i,k} = 1}$, and $\sigma(Y_{i,k}^{\operatorname{diff}}) = Y_{i,\sigma_i(k)}^{\operatorname{diff}}$. Following this notation, by Theorem \ref{theo:key_perm}, it suffices to show that for any allowed permutation $\sigma$,
\begin{equation}
\label{eqn:goal_theorem2}
\sigma(Y^{\operatorname{diff}}_{\mathcal{I}_1,1:K})
\mid W_{1:n,1:K}, X_{1:n}, \mathcal{I}_m \stackrel{d}{=}
Y^{\operatorname{diff}}_{\mathcal{I}_1,1:K}
\mid W_{1:n,1:K}, X_{1:n}, \mathcal{I}_m.
\end{equation}

Under Assumptions \ref{assu:no_temp_interf} - \ref{assu:time_fixed_effect} and Hypothesis \ref{hypo:no_cu_interference}, following \eqref{eqn:simple_potential_outcome_hypo2}, we can write $Y_{i,k}(w) = \alpha_i(w) + u_k + \epsilon_{i,k}(w)$. Therefore, for any $i \in \mathcal{I}_1$ and $k \in S_i$, we have that $Y_{i,k} = Y_{i,k}(1) = \alpha_i(1) + u_k + \epsilon_{i,k}(1)$. At the same time, for the matched unit of $i$, we have $W_{m(i),k} = 0$, and thus $Y_{m(i),k} = Y_{m(i),k}(0) = \alpha_{m(i)}(0) + u_k + \epsilon_{m(i),k}(0)$. The difference of the two satisfies 
\begin{equation}
\begin{split}
    Y^{\operatorname{diff}}_{i,k} = Y_{i,k} - Y_{m(i),k}
    &= \alpha_i(1) + u_k + \epsilon_{i,k}(1) - \p{\alpha_{m(i)}(0) + u_k + \epsilon_{m(i),k}(0)}\\
    &= \alpha_i(1) + \epsilon_{i,k}(1) - \alpha_{m(i)}(0) + \epsilon_{m(i),k}(0).
\end{split}
\end{equation}
Under Assumption \ref{assu:time_fixed_effect}, we have that
\begin{equation}
\begin{split}
    &\p{\alpha_i(1) + \epsilon_{i,k}(1) - \alpha_{m(i)}(0) + \epsilon_{m(i),k}(0)}
    \mid W_{1:n,1:K}, X_{1:n}, \mathcal{I}_m, \alpha_{1:n}\\
    & \qquad \qquad \qquad \qquad \stackrel{d}{=}
    \p{\alpha_i(1) + \epsilon_{i,\sigma_i(k)}(1) - \alpha_{m(i)}(0) + \epsilon_{m(i),\sigma_i(k)}(0)}
    \mid W_{1:n, 1:K}, X_{1:n}, \mathcal{I}_m, \alpha_{1:n}
\end{split}
\end{equation}
\sloppy{for any permutation $\sigma_i$ of $S_i$, because the errors $\epsilon_{i,k}$ and $\epsilon_{i,\sigma_i(k)}$ are i.i.d conditioning on $W_{1:n, 1:K}, X_{1:n}$ and $\alpha_{1:n}$ (and same for $\epsilon_{m(i),k}$ and $\epsilon_{m(i),\sigma_i(k)}$). In addition, since all the errors $\epsilon_{i,k}$'s are independent conditioning on $W_{1:n, 1:K}, X_{1:n}$ and $\alpha_{1:n}$, we have that}
\begin{equation}
\begin{split}
    &\p{\alpha_i(1) + \epsilon_{i,k}(1) - \alpha_{m(i)}(0) + \epsilon_{m(i),k}(0)}_{i \in \mathcal{I}_1}
    \mid W_{1:n,1:K}, X_{1:n}, \mathcal{I}_m, \alpha_{1:n}\\
    & \qquad \qquad  \stackrel{d}{=}
    \p{\alpha_i(1) + \epsilon_{i,\sigma_i(k)}(1) - \alpha_{m(i)}(0) + \epsilon_{m(i),\sigma_i(k)}(0)}_{i \in \mathcal{I}_1}
    \mid W_{1:n,1:K}, X_{1:n}, \mathcal{I}_m,\alpha_{1:n}.
\end{split}
\end{equation}
Rewriting the above, we get
\begin{equation}
Y^{\operatorname{diff}}_{\mathcal{I}_1, 1:K}, \alpha_{1:n}
\mid W_{1:n,1:K}, X_{1:n}, \mathcal{I}_m \stackrel{d}{=}
\sigma(Y^{\operatorname{diff}}_{\mathcal{I}_1, 1:K})
\mid W_{1:n,1:K}, X_{1:n}, \mathcal{I}_m, \alpha_{1:n},
\end{equation}
which further implies \eqref{eqn:goal_theorem2} and hence gives the desired result. 

\end{proof}

\section{Simulations}
\label{section:simulation}

In this section, we focus on a form of network interference. Specifically, we use a real-life social network to describe social interactions among units. We generate outcomes with some magnitude of network interference and evaluate our methods based on these generated outcomes. Our simulations can be viewed as semi-synthetic experiments---we use a real-life network, but we generate outcomes according to some model.

We consider the Swarthmore network in the Facebook 100 dataset \citep{TRAUD20124165}. All networks in this dataset are complete online friendship networks for one hundred colleges and universities collected from a single-day snapshot of Facebook in September 2005. Here we focus on the Swarthmore college network in our simulation. To make the social network connected, we extract the largest connected component of the Swarthmore network. To summarize, the network we use is of size 1657 with 61049 edges. The diameter of the network is 6 and the average pairwise distance is 2.32.

Throughout this section, we assume that we have access to the data of three randomized experiments. We take treatment probabilities $\pi_1 = 10\%$,  $\pi_2 = 25\%$ and $\pi_3 = 50\%$. 
In the following simulation studies, we consider level of significance $\alpha = 0.05$. Every dot on each plot is an average over 500 replications. We take $B = 200$. 

\subsection{Under general assumptions}
\label{section:simulation_general}

We compare the power of the tests given in Algorithms~\ref{alg:test_one_exp}, \ref{alg:test_two_exp} and \ref{alg:test_three_exp}. We run Algorithm \ref{alg:test_three_exp} using all three experiments, run Algorithm~\ref{alg:test_two_exp} using the second and the third experiments, and run Algorithm~\ref{alg:test_one_exp} using the third experiment, i.e., we always use experiments with the largest treatment probabilities. We discuss the choice of test statistics in Appendix \ref{section:simulation_detal}. In Figure \ref{fig:vertical_linear}, we assume a linear model of the outcome $Y$; in Figure \ref{fig:vertical_nonlinear}, we assume a nonlinear model. The details of the generating model can also be found in Appendix~\ref{section:simulation_detal}. 

\begin{figure}
\begin{subfigure}[b]{\textwidth}
%\raggedleft
\caption{Outcome $Y$ follows a linear model.}
\includegraphics[width = \textwidth]{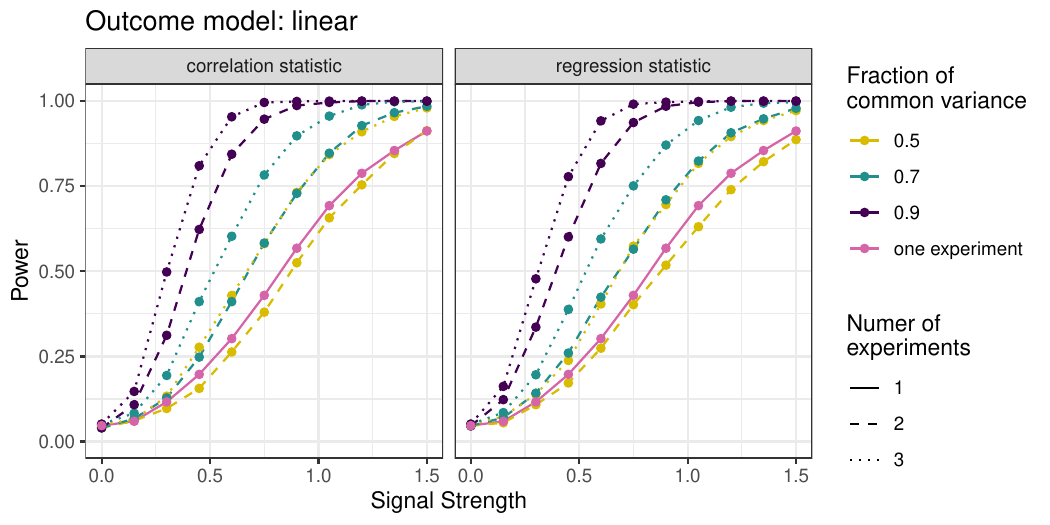}
\label{fig:vertical_linear}
\end{subfigure}
    \hfill
\begin{subfigure}[b]{\textwidth}
%\raggedleft
\caption{Outcome $Y$ follows a nonlinear model.}
\includegraphics[width = \textwidth]{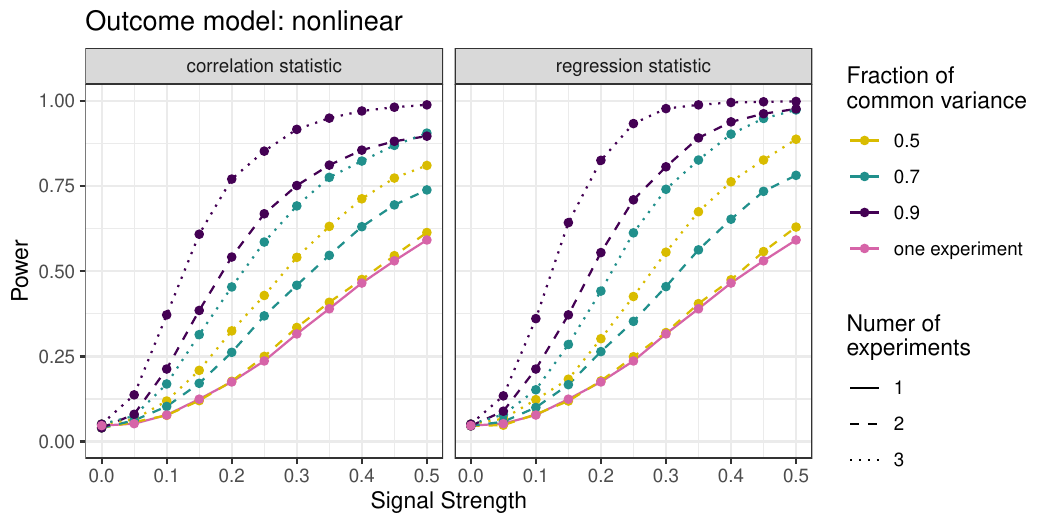}
\label{fig:vertical_nonlinear}
\end{subfigure}

\caption{Power of Algorithms~\ref{alg:test_one_exp}, \ref{alg:test_two_exp} and \ref{alg:test_three_exp}. }
\end{figure}

In Figures \ref{fig:vertical_linear} and \ref{fig:vertical_nonlinear}, we plot the power of the testing algorithms~\ref{alg:test_one_exp}, \ref{alg:test_two_exp} and \ref{alg:test_three_exp} at different levels of interference effects (signal strengths). In the figures, the fraction of common variance controls the correlation of the individual outcomes across experiments.

We observe from Figures \ref{fig:vertical_linear} and \ref{fig:vertical_nonlinear} that utilizing more experiments helps our algorithms become more powerful, especially when the fraction of common variance is high. As discussed in Section \ref{section:related_work}, our work is the first to consider testing interference with multiple randomized experiments. Therefore, we can treat the algorithm utilizing one experiment as the \textit{baseline method} that represents the state-of-the-art. Our algorithms appear to have a clear advantage over the baseline in terms of the power.

We also find that the regression statistic performs better than the correlation statistic, because the regression step helps reduce variance caused by the observed covariates.

\subsection{Time fixed effect model}
\label{section:simulation_time_fixed}

We compare the power of the tests given in Algorithms~\ref{alg:test_three_exp} and \ref{alg:hori_perm_multi}. We run both algorithms using all three experiments. We use a regression test statistic in both algorithms. We discuss the choice of test statistics and matching algorithms in Appendix \ref{section:simulation_detal}. In Figure \ref{fig:horizontal_linear}, we assume a linear model of the outcome $Y$, whereas in Figure \ref{fig:horizontal_nonlinear}, we assume a nonlinear model. The details of the generating model can also be found in Appendix~\ref{section:simulation_detal}. 

\begin{figure}
\begin{subfigure}[b]{\textwidth}
%\raggedleft
\caption{Outcome $Y$ follows a linear model.}
\includegraphics[width = \textwidth]{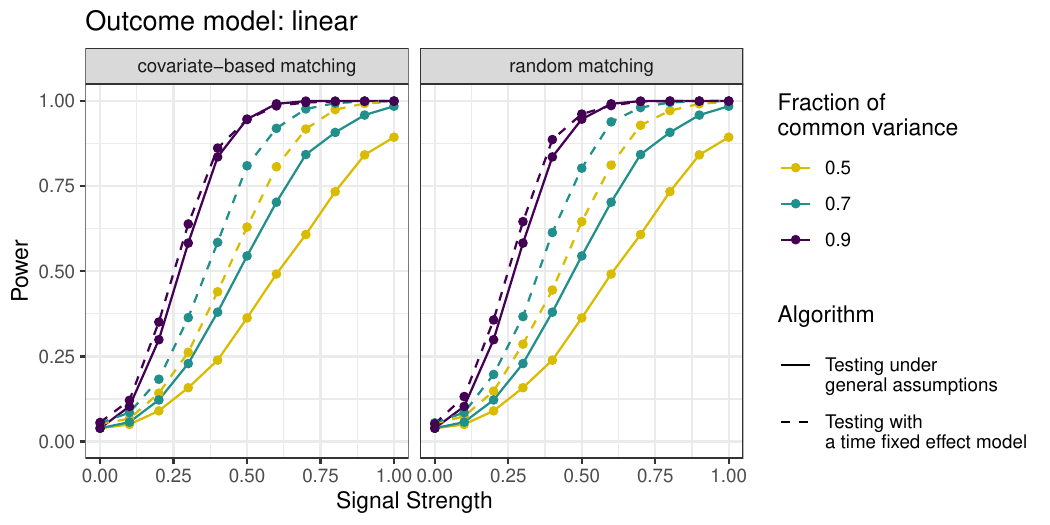}
\label{fig:horizontal_linear}
\end{subfigure}
    \hfill
\begin{subfigure}[b]{\textwidth}
%\raggedleft
\caption{Outcome $Y$ follows a nonlinear model.}
\includegraphics[width = \textwidth]{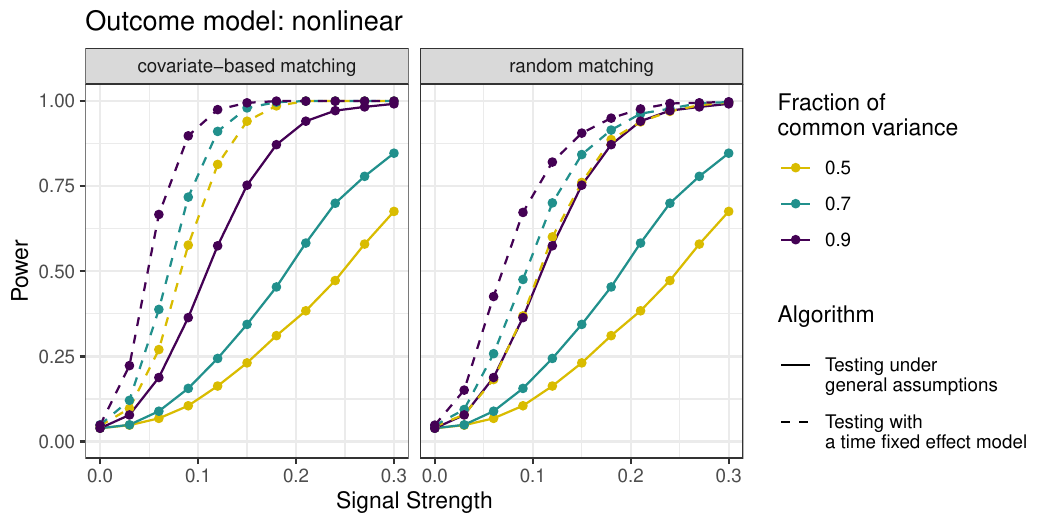}
\label{fig:horizontal_nonlinear}
\end{subfigure}

\caption{Power of Algorithms \ref{alg:test_three_exp} and \ref{alg:hori_perm_multi}. }
\end{figure}

In Figures \ref{fig:horizontal_linear} and \ref{fig:horizontal_nonlinear}, we plot the power of the testing algorithms~\ref{alg:test_three_exp} and \ref{alg:hori_perm_multi} at different levels of interference effects (signal strengths). Algorithm \ref{alg:hori_perm_multi} (testing with
a time fixed effect model) appears more powerful than Algorithm \ref{alg:test_three_exp} (testing under general assumptions). To understand this phenomenon, we recall that Algorithm~\ref{alg:test_three_exp} permutes data across experiments, whereas Algorithm~\ref{alg:hori_perm_multi}
permutes data across units. Due to the nature of A/B tests, there is more variability in treatment allocation across experiments than across units. For example, assume that all units have around $n_{\operatorname{ngb}}$ neighbors in the social network. Looking at the fraction of neighbors in the treatment group, we find that the variation of this quantity across units is of scale $1/\sqrt{n_{\operatorname{ngb}}}$, whereas the variation of this quantity across experiments is of constant scale. By permuting over data points that are more different, Algorithm~\ref{alg:hori_perm_multi} gains extra power. 

Recall that there is a matching step in Algorithm \ref{alg:hori_perm_multi}. We find from Figure \ref{fig:horizontal_linear} and \ref{fig:horizontal_nonlinear} that covariate-based matching outperforms random matching, especially under a nonlinear outcome model. In a linear model, the regression step has already removed almost all of the variance caused by observed covariates. In a nonlinear model, nevertheless, the regression step cannot fully remove all variance and the matching step can help further reduce variance. 

\section{Applications}
\label{section:real_data}

In this section, we illustrate how the proposed procedure has been successfully implemented at LinkedIn as an add-on to their experimentation toolkit. Like other firms in the technology sector such as Google and Meta, LinkedIn makes business decisions in a data-driven manner and has a culture to “test everything”. To support the needs to run concurrent A/B tests at scale, LinkedIn built an in-house experimentation platform, called T-REX (Targeting, Ramping, and Experimentation), which provides end-to-end experimentation supports \citep{xu2015infrastructure, ivanuik2020}. Regardless of the application, T-REX implements simple Bernoulli randomization and relies on $t$-test for readout without taking into account potential interactions among experimental units.

This becomes a major limitation for experimentation in a marketplace environment, including the ads marketplace, where units on either side of the marketplace (advertisers and ad viewers) can interfere with each other \citep{basse2016randomization, pouget2019variance, liu2021trustworthy, johari2022experimental}. For example, ad campaigns that share the  targeting audiences interfere with each other by competing in auctions for ad slots; different ad viewers with similar attributes are connected through the finite budget of certain ad campaigns. To remove bias in experiments caused by interference, LinkedIn has implemented the Budget-split platform on top of T-REX for experimentation in their ads marketplace \citep{liu2021trustworthy}. 

However, since Budget-split uses two halves of the marketplace to simulate the counterfactuals under different treatment variants, it does not support the classic factorial design. Under the current implementation, the platform only runs one experiment at a time, which is much smaller than the total number of experiments they need to run. This limitation in Budget-split capacity severely delays innovation: teams need to wait for weeks for a Budget-split slot in order to get an accurate measurement of their feature ramp before product launch. Nevertheless, not all ramps suffer from unit interaction, even in the ads marketplace setting. Running Budget-split experiments with negligible interference incurs a huge opportunity cost. Ideally, the Budget-split platform wants to prioritize tests that are impacted the most by the interference effects.

%\begin{figure}[t]
%    \centering
%    \includegraphics[width = 0.9\textwidth]{plots/app_design.png}
%    \caption{Interference detector in LinkedIn's ads experimentation system.}
%    \label{fig:app_design}
%\end{figure}

At LinkedIn, all feature launches start with small percentage ramps for risk mitigation and gradually increase the treatment percentage (i.e., 1\%, 5\%, 10\%, 25\%) before reaching the iteration for treatment effect measurement (50\%) \citep{xu2018sqr, mao2021quantifying}. Specifically, Budget-split amounts to a 50\% ramp on the viewers’ side. This increasing allocation scheme provides us information to detect potential interference. With the algorithms proposed in this paper, we implemented a screening step for each feature after the 25\% iteration. The experiments are then ranked by the $p$-value in the interference test to determine their priority on the Budget-split platform.

It is important to note that the screening module was designed as an add-on to the system without touching LinkedIn’s existing experimentation solution such as T-REX. %Figure~\ref{fig:app_design} illustrates its niche in the feature release cycle. 
By default, the interference detector only requires experimentation data in two previous iterations and runs Algorithm~\ref{alg:hori_perm}. Users have the option to provide additional network information that characterizes the potential interference mechanism among units and run other algorithms in this paper. Because of this standalone nature, a similar interference detector can be readily added to any existing experimentation platforms to trigger alerts when interference might cause a problem.

\begin{figure}[t]
    \centering
    \includegraphics[width =0.49\textwidth]{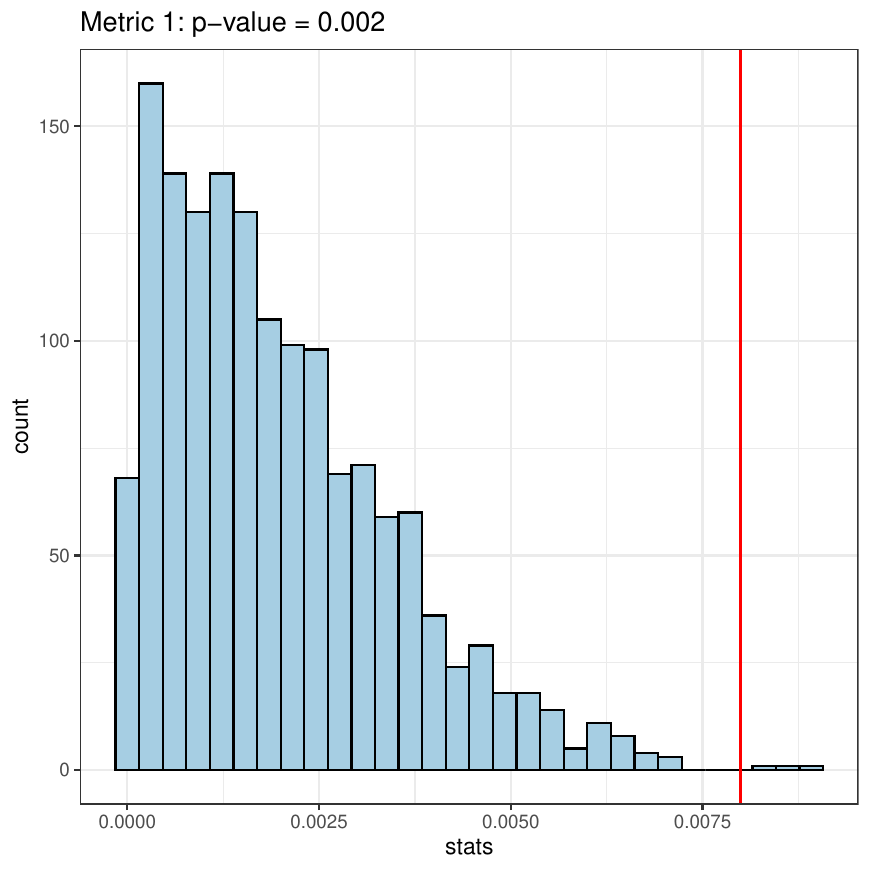}
    \includegraphics[width = 0.49\textwidth]{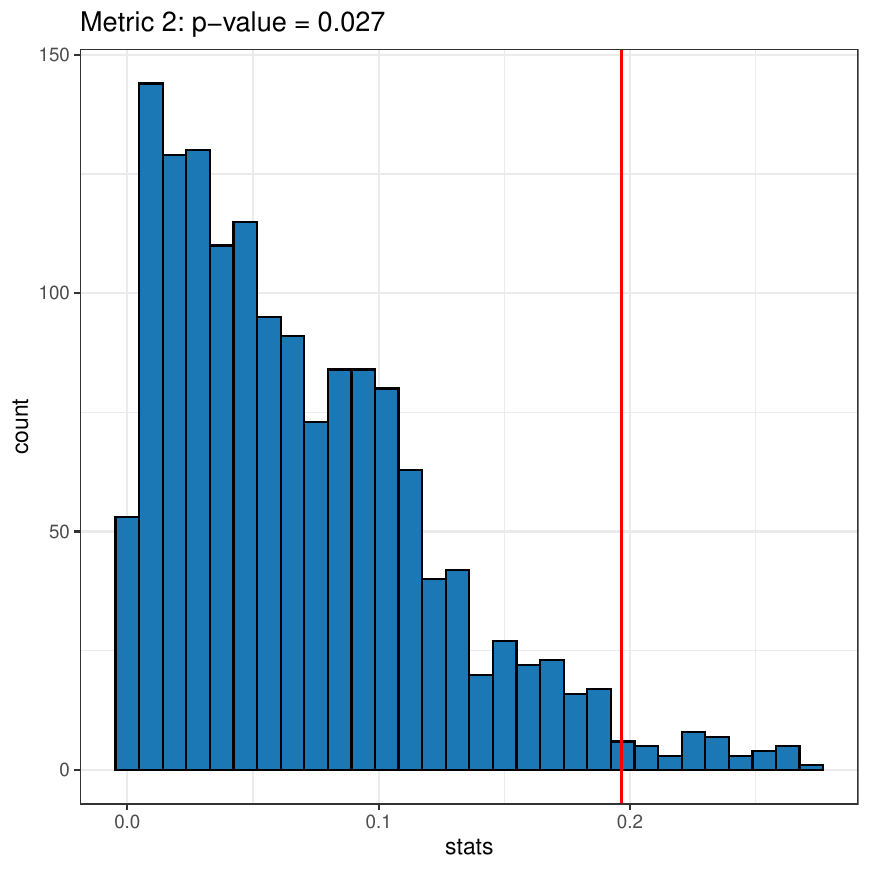}
    \caption{Example experiment: Test statistics and $p$-values from the permutation test. Results on two metrics are shown.}
    \label{fig:app_hist_1}
\end{figure}

As an illustration, we consider an online controlled experiment implemented by LinkedIn. The treatment in this experiment corresponds to a new feature that improves the quality of LinkedIn members' attribute for ads targeting. We run a series of experiments with increasing allocation with the members as the randomization units. Interference effect is expected in these experiments: when the allocation percentage is small, only a small set of members have the updated attributes, making them easier to be targeted by ad campaigns. Thus, when comparing metrics such as total ad impressions, these members tend to have larger average results than members in the control group. When the treatment allocation increases, more members get the improved attributes. Since the total ad budget does not increase much, the average difference between treatment and control units becomes smaller. Figure~\ref{fig:example_DIM} shows the average differences between treatment and control units in the experiment series. Figure~\ref{fig:app_hist_1} shows the output from the interference detector after running Algorithm~\ref{alg:hori_perm} based on the 10\% and 25\% iterations with respect to two different metrics. The $p$-values of the permutation test confirm the strong interference effects in these experiments.

%we consider the experiment in Figure~\ref{fig:example_DIM}. Figure~\ref{fig:app_hist_1} shows the output from the interference detector after running Algorithm~\ref{alg:hori_perm} based on the 10\% and 25\% iterations with respect to two different metrics. The $p$-values of the permutation test suggests that there are strong interference effects in this experiment. It turns out that the randomization units in this experiment are LinkedIn members. The feature being tested is introduced to improve the quality of LinkedIn members' attribute for ads targeting. When the allocation percentage is small, only a small set of members have the updated attributes, making them easier to be targeted by ad campaigns. Thus when comparing metrics such as total ad impressions, these member tends to have larger average results than members in the control group. When the treatment allocation increase, more members get the improved attributes. Since the total ad budget does not increase much, the average difference between treatment and control units becomes smaller. 

\section{Discussion}
\paragraph{Missingness.} In this paper, we make the assumption that the dataset is complete. A natural future direction of work is to extend the current methods to scenarios with missing data. It is not hard to show that if the data is missing completely at random (MCAR), then the proposed testing procedures are still valid. When MCAR is unrealistic, it will be interesting to study whether our methods can still be applied under certain conditions. In practice, experimenters need to carefully examine the possible causes and consequences of missingness and make decisions correspondingly. 

\paragraph{Selective inference.}
We propose to use our testing procedure as a screening step for A/B testing: if the test suggests that no interference exists, then the experimenter can proceed with classical causal inference analysis. Strictly speaking, the data is used twice here---in the screening step and in the follow-up analysis.  It would be of interest to understand the impact of the screening step on the follow-up analysis, and to develop valid statistical inference methods conditioning on the result of the screening step. 

\paragraph{Sequential Testing.}
Another question left open by this paper is whether the proposed methods can be extended to the sequential testing setting. Our current procedure fixes the number of experiments a priori and constructs a single $p$-value from the permutation test. In real life, the treatment probability increases gradually, and it would be of practical interest to end the experiment early as soon as we detect any interference. In that scenario, we need to take into account the randomness in stopping time and construct always valid $p$-values \citep{johari2017}.

\section*{Acknowledgements}
S.L. was supported by NIH/NIDA grant P50 DA054039 and NIH/NIDCR grant UH3 DE028723.
The authors are grateful to Iavor Bojinov, Ari Boyarsky, Justin Dyer, Guido Imbens, Jess Jackson, Hongseok Namkoong, Jean Pouget-Abadie, Johan Ugander and Stefan Wager for their constructive feedback and insightful comments.

\bibliographystyle{plainnat}
\bibliography{references}

\begin{thebibliography}{65}
\providecommand{\natexlab}[1]{#1}
\providecommand{\url}[1]{\texttt{#1}}
\expandafter\ifx\csname urlstyle\endcsname\relax
  \providecommand{\doi}[1]{doi: #1}\else
  \providecommand{\doi}{doi: \begingroup \urlstyle{rm}\Url}\fi

\bibitem[Angrist and Pischke(2009)]{angrist2009mostly}
Joshua~D Angrist and J{\"o}rn-Steffen Pischke.
\newblock \emph{Mostly harmless econometrics: An empiricist's companion}.
\newblock Princeton university press, 2009.

\bibitem[Aronow(2012)]{aronow2012}
Peter~M. Aronow.
\newblock A general method for detecting interference between units in
  randomized experiments.
\newblock \emph{Sociological Methods \& Research}, 41\penalty0 (1):\penalty0
  3--16, 2012.

\bibitem[Aronow and Samii(2017)]{arownow2017}
Peter~M. Aronow and Cyrus Samii.
\newblock Estimating average causal effects under general interference, with
  application to a social network experiment.
\newblock \emph{The Annals of Applied Statistics}, 11\penalty0 (4):\penalty0
  1912--1947, 2017.

\bibitem[Athey et~al.(2018)Athey, Eckles, and Imbens]{athey2018exact}
Susan Athey, Dean Eckles, and Guido~W Imbens.
\newblock Exact $p$-values for network interference.
\newblock \emph{Journal of the American Statistical Association}, 113\penalty0
  (521):\penalty0 230--240, 2018.

\bibitem[Bajari et~al.(2021)Bajari, Burdick, Imbens, Masoero, McQueen,
  Richardson, and Rosen]{bajari2021multiple}
Patrick Bajari, Brian Burdick, Guido~W Imbens, Lorenzo Masoero, James McQueen,
  Thomas Richardson, and Ido~M Rosen.
\newblock Multiple randomization designs.
\newblock \emph{arXiv preprint arXiv:2112.13495}, 2021.

\bibitem[Bakshy et~al.(2014)Bakshy, Eckles, and Bernstein]{bakshy2014designing}
Eytan Bakshy, Dean Eckles, and Michael~S Bernstein.
\newblock Designing and deploying online field experiments.
\newblock In \emph{Proceedings of the 23rd international conference on World
  wide web}, pages 283--292, 2014.

\bibitem[Basse et~al.(2019)Basse, Feller, and Toulis]{basse2019}
G~W Basse, A~Feller, and P~Toulis.
\newblock {Randomization tests of causal effects under interference}.
\newblock \emph{Biometrika}, 106\penalty0 (2):\penalty0 487--494, 02 2019.

\bibitem[Basse and Feller(2018)]{basse2018}
Guillaume Basse and Avi Feller.
\newblock Analyzing two-stage experiments in the presence of interference.
\newblock \emph{Journal of the American Statistical Association}, 113\penalty0
  (521):\penalty0 41--55, 2018.
\newblock \doi{10.1080/01621459.2017.1323641}.

\bibitem[Basse and Airoldi(2018)]{basse_airoldi_2018}
Guillaume~W. Basse and Edoardo~M. Airoldi.
\newblock Limitations of design-based causal inference and a/b testing under
  arbitrary and network interference.
\newblock \emph{Sociological Methodology}, 48\penalty0 (1):\penalty0 136--151,
  2018.
\newblock \doi{10.1177/0081175018782569}.

\bibitem[Basse et~al.(2016)Basse, Soufiani, and
  Lambert]{basse2016randomization}
Guillaume~W Basse, Hossein~Azari Soufiani, and Diane Lambert.
\newblock Randomization and the pernicious effects of limited budgets on
  auction experiments.
\newblock In \emph{Artificial Intelligence and Statistics}, pages 1412--1420.
  PMLR, 2016.

\bibitem[Bertrand et~al.(2004)Bertrand, Duflo, and
  Mullainathan]{bertrand2004much}
Marianne Bertrand, Esther Duflo, and Sendhil Mullainathan.
\newblock How much should we trust differences-in-differences estimates?
\newblock \emph{The Quarterly journal of economics}, 119\penalty0 (1):\penalty0
  249--275, 2004.

\bibitem[Bhattacharya et~al.(2020)Bhattacharya, Malinsky, and
  Shpitser]{bhattacharya2020causal}
Rohit Bhattacharya, Daniel Malinsky, and Ilya Shpitser.
\newblock Causal inference under interference and network uncertainty.
\newblock In \emph{Uncertainty in Artificial Intelligence}, pages 1028--1038.
  PMLR, 2020.

\bibitem[Bojinov et~al.(2021)Bojinov, Rambachan, and Shephard]{bojinov2021}
Iavor Bojinov, Ashesh Rambachan, and Neil Shephard.
\newblock Panel experiments and dynamic causal effects: A finite population
  perspective.
\newblock \emph{Quantitative Economics}, 12\penalty0 (4):\penalty0 1171--1196,
  2021.

\bibitem[Bowers et~al.(2013)Bowers, Fredrickson, and
  Panagopoulos]{bowers_fredrickson_panagopoulos_2013}
Jake Bowers, Mark~M. Fredrickson, and Costas Panagopoulos.
\newblock Reasoning about interference between units: A general framework.
\newblock \emph{Political Analysis}, 21\penalty0 (1):\penalty0 97–124, 2013.
\newblock \doi{10.1093/pan/mps038}.

\bibitem[Cochran and Rubin(1973)]{cochran1973controlling}
William~G Cochran and Donald~B Rubin.
\newblock Controlling bias in observational studies: A review.
\newblock \emph{Sankhy{\=a}: The Indian Journal of Statistics, Series A}, pages
  417--446, 1973.

\bibitem[Cortez et~al.(2022)Cortez, Eichhorn, and Yu]{cortez2022graph}
Mayleen Cortez, Matthew Eichhorn, and Christina~Lee Yu.
\newblock Graph agnostic estimators with staggered rollout designs under
  network interference.
\newblock \emph{arXiv preprint arXiv:2205.14552}, 2022.

\bibitem[Eckles et~al.(2017)Eckles, Karrer, and
  Ugander]{EcklesKarrerUgander+2017}
Dean Eckles, Brian Karrer, and Johan Ugander.
\newblock Design and analysis of experiments in networks: Reducing bias from
  interference.
\newblock \emph{Journal of Causal Inference}, 5\penalty0 (1):\penalty0
  20150021, 2017.

\bibitem[Fisher(1925)]{fisher1925statistical}
Ronald~Aylmer Fisher.
\newblock \emph{Statistical Methods for Research Workers}.
\newblock Number~3. Oliver and Boyd, 1925.

\bibitem[Fradkin(2019)]{fradkin2019simulation}
Andrey Fradkin.
\newblock A simulation approach to designing digital matching platforms.
\newblock \emph{Boston University Questrom School of Business Research Paper
  Forthcoming}, 2019.

\bibitem[Fujikoshi(1993)]{fujikoshi1993two}
Yasunori Fujikoshi.
\newblock Two-way anova models with unbalanced data.
\newblock \emph{Discrete Mathematics}, 116\penalty0 (1-3):\penalty0 315--334,
  1993.

\bibitem[Han et~al.(2021)Han, Bojinov, and Basse]{Han2021}
Kevin~Wu Han, Iavor Bojinov, and Guillaume Basse.
\newblock Population interference in panel experiments, 2021.
\newblock URL \url{https://arxiv.org/abs/2103.00553}.

\bibitem[Hansen and Klopfer(2006)]{hansen2006optimal}
Ben~B Hansen and Stephanie~Olsen Klopfer.
\newblock Optimal full matching and related designs via network flows.
\newblock \emph{Journal of computational and Graphical Statistics}, 15\penalty0
  (3):\penalty0 609--627, 2006.

\bibitem[Hemerik and Goeman(2018{\natexlab{a}})]{hemerik2018exact}
Jesse Hemerik and Jelle Goeman.
\newblock Exact testing with random permutations.
\newblock \emph{Test}, 27\penalty0 (4):\penalty0 811--825, 2018{\natexlab{a}}.

\bibitem[Hemerik and Goeman(2018{\natexlab{b}})]{hemerik2018false}
Jesse Hemerik and Jelle~J Goeman.
\newblock False discovery proportion estimation by permutations: confidence for
  significance analysis of microarrays.
\newblock \emph{Journal of the Royal Statistical Society: Series B (Statistical
  Methodology)}, 80\penalty0 (1):\penalty0 137--155, 2018{\natexlab{b}}.

\bibitem[Holtz et~al.(2020)Holtz, Lobel, Liskovich, and
  Aral]{holtz2020reducing}
David Holtz, Ruben Lobel, Inessa Liskovich, and Sinan Aral.
\newblock Reducing interference bias in online marketplace pricing experiments.
\newblock \emph{arXiv preprint arXiv:2004.12489}, 2020.

\bibitem[Hu et~al.(2022)Hu, Li, and Wager]{hu2022}
Yuchen Hu, Shuangning Li, and Stefan Wager.
\newblock {Average direct and indirect causal effects under interference}.
\newblock \emph{Biometrika}, 02 2022.
\newblock ISSN 1464-3510.
\newblock \doi{10.1093/biomet/asac008}.
\newblock URL \url{https://doi.org/10.1093/biomet/asac008}.
\newblock asac008.

\bibitem[Hudgens and Halloran(2008)]{hudgens_2008}
Michael~G Hudgens and M.~Elizabeth Halloran.
\newblock Toward causal inference with interference.
\newblock \emph{Journal of the American Statistical Association}, 103\penalty0
  (482):\penalty0 832--842, 2008.

\bibitem[Imbens and Rubin(2015)]{imbens_rubin_2015}
Guido~W. Imbens and Donald~B. Rubin.
\newblock \emph{Causal Inference for Statistics, Social, and Biomedical
  Sciences: An Introduction}.
\newblock Cambridge University Press, 2015.
\newblock \doi{10.1017/CBO9781139025751}.

\bibitem[Ivaniuk(2020)]{ivanuik2020}
Alexander Ivaniuk.
\newblock Our evolution towards t-rex: The prehistory of experimentation
  infrastructure at linkedin.
\newblock \emph{LinkedIn Engineering Blog}, 2020.

\bibitem[Johari et~al.(2017)Johari, Koomen, Pekelis, and Walsh]{johari2017}
Ramesh Johari, Pete Koomen, Leonid Pekelis, and David Walsh.
\newblock Peeking at {A/B} tests: Why it matters, and what to do about it.
\newblock In \emph{Proceedings of the 23rd ACM SIGKDD International Conference
  on Knowledge Discovery and Data Mining}, KDD '17, page 1517–1525, New York,
  NY, USA, 2017. Association for Computing Machinery.

\bibitem[Johari et~al.(2022)Johari, Li, Liskovich, and
  Weintraub]{johari2022experimental}
Ramesh Johari, Hannah Li, Inessa Liskovich, and Gabriel~Y Weintraub.
\newblock Experimental design in two-sided platforms: An analysis of bias.
\newblock \emph{Management Science}, 2022.

\bibitem[Kohavi et~al.(2013)Kohavi, Deng, Frasca, Walker, Xu, and
  Pohlmann]{kohavi_deng_2013}
Ron Kohavi, Alex Deng, Brian Frasca, Toby Walker, Ya~Xu, and Nils Pohlmann.
\newblock Online controlled experiments at large scale.
\newblock In \emph{Proceedings of the 19th ACM SIGKDD International Conference
  on Knowledge Discovery and Data Mining}, KDD '13, page 1168–1176, New York,
  NY, USA, 2013. Association for Computing Machinery.
\newblock ISBN 9781450321747.
\newblock \doi{10.1145/2487575.2488217}.
\newblock URL \url{https://doi.org/10.1145/2487575.2488217}.

\bibitem[Kohavi et~al.(2020)Kohavi, Tang, and Xu]{kohavi_tang_xu_2020}
Ron Kohavi, Diane Tang, and Ya~Xu.
\newblock \emph{Trustworthy Online Controlled Experiments: A Practical Guide to
  A/B Testing}.
\newblock Cambridge University Press, 2020.
\newblock \doi{10.1017/9781108653985}.

\bibitem[Leung(2020)]{leung2020}
Michael~P. Leung.
\newblock Treatment and spillover effects under network interference.
\newblock \emph{The Review of Economics and Statistics}, 102\penalty0
  (2):\penalty0 368--380, 05 2020.
\newblock ISSN 0034-6535.
\newblock \doi{10.1162/rest_a_00818}.

\bibitem[Li et~al.(2022)Li, Zhao, Johari, and Weintraub]{li2022interference}
Hannah Li, Geng Zhao, Ramesh Johari, and Gabriel~Y Weintraub.
\newblock Interference, bias, and variance in two-sided marketplace
  experimentation: Guidance for platforms.
\newblock In \emph{Proceedings of the ACM Web Conference 2022}, pages 182--192,
  2022.

\bibitem[Li and Wager(2022)]{li2022}
Shuangning Li and Stefan Wager.
\newblock {Random graph asymptotics for treatment effect estimation under
  network interference}.
\newblock \emph{The Annals of Statistics}, 50\penalty0 (4):\penalty0 2334 --
  2358, 2022.
\newblock \doi{10.1214/22-AOS2191}.
\newblock URL \url{https://doi.org/10.1214/22-AOS2191}.

\bibitem[Liu et~al.(2021)Liu, Mao, and Kang]{liu2021trustworthy}
Min Liu, Jialiang Mao, and Kang Kang.
\newblock Trustworthy and powerful online marketplace experimentation with
  budget-split design.
\newblock In \emph{Proceedings of the 27th ACM SIGKDD Conference on Knowledge
  Discovery \& Data Mining}, pages 3319--3329, 2021.

\bibitem[Mahmood(2018)]{mahmood2018performance}
Sharif Mahmood.
\newblock The performance of largest caliper matching: A monte carlo simulation
  approach.
\newblock \emph{arXiv preprint arXiv:1806.02149}, 2018.

\bibitem[Mao and Bojinov(2021)]{mao2021quantifying}
Jialiang Mao and Iavor Bojinov.
\newblock Quantifying the value of iterative experimentation.
\newblock \emph{arXiv preprint arXiv:2111.02334}, 2021.

\bibitem[Pouget-Abadie et~al.(2019{\natexlab{a}})Pouget-Abadie, Aydin, Schudy,
  Brodersen, and Mirrokni]{pouget2019variance}
Jean Pouget-Abadie, Kevin Aydin, Warren Schudy, Kay Brodersen, and Vahab
  Mirrokni.
\newblock Variance reduction in bipartite experiments through correlation
  clustering.
\newblock \emph{Advances in Neural Information Processing Systems}, 32,
  2019{\natexlab{a}}.

\bibitem[Pouget-Abadie et~al.(2019{\natexlab{b}})Pouget-Abadie, Saint-Jacques,
  Saveski, Duan, Ghosh, Xu, and Airoldi]{pouget2019testing}
Jean Pouget-Abadie, Guillaume Saint-Jacques, Martin Saveski, Weitao Duan,
  S~Ghosh, Y~Xu, and Edoardo~M Airoldi.
\newblock Testing for arbitrary interference on experimentation platforms.
\newblock \emph{Biometrika}, 106\penalty0 (4):\penalty0 929--940,
  2019{\natexlab{b}}.

\bibitem[Puelz et~al.(2022)Puelz, Basse, Feller, and Toulis]{puelz2022}
David Puelz, Guillaume Basse, Avi Feller, and Panos Toulis.
\newblock {A graph‐theoretic approach to randomization tests of causal
  effects under general interference}.
\newblock \emph{Journal of the Royal Statistical Society Series B}, 84\penalty0
  (1):\penalty0 174--204, February 2022.

\bibitem[Rosenbaum(1989)]{rosenbaum1989optimal}
Paul~R Rosenbaum.
\newblock Optimal matching for observational studies.
\newblock \emph{Journal of the American Statistical Association}, 84\penalty0
  (408):\penalty0 1024--1032, 1989.

\bibitem[Rosenbaum and Rubin(1985)]{rosenbaum1985constructing}
Paul~R Rosenbaum and Donald~B Rubin.
\newblock Constructing a control group using multivariate matched sampling
  methods that incorporate the propensity score.
\newblock \emph{The American Statistician}, 39\penalty0 (1):\penalty0 33--38,
  1985.

\bibitem[Rubin(1973)]{rubin1973matching}
Donald~B Rubin.
\newblock Matching to remove bias in observational studies.
\newblock \emph{Biometrics}, pages 159--183, 1973.

\bibitem[Rubin(1980)]{rubin1980bias}
Donald~B Rubin.
\newblock Bias reduction using mahalanobis-metric matching.
\newblock \emph{Biometrics}, pages 293--298, 1980.

\bibitem[Saveski et~al.(2017)Saveski, Pouget-Abadie, Saint-Jacques, Duan,
  Ghosh, Xu, and Airoldi]{saveski2017detecting}
Martin Saveski, Jean Pouget-Abadie, Guillaume Saint-Jacques, Weitao Duan,
  Souvik Ghosh, Ya~Xu, and Edoardo~M Airoldi.
\newblock Detecting network effects: Randomizing over randomized experiments.
\newblock In \emph{Proceedings of the 23rd ACM SIGKDD international conference
  on knowledge discovery and data mining}, pages 1027--1035, 2017.

\bibitem[S{\"a}vje et~al.(2021)S{\"a}vje, Aronow, and Hudgens]{savje2021}
Fredrik S{\"a}vje, Peter~M. Aronow, and Michael~G. Hudgens.
\newblock {Average treatment effects in the presence of unknown interference}.
\newblock \emph{The Annals of Statistics}, 49\penalty0 (2):\penalty0 673 --
  701, 2021.
\newblock \doi{10.1214/20-AOS1973}.
\newblock URL \url{https://doi.org/10.1214/20-AOS1973}.

\bibitem[Sekhon(2008)]{sekhon2008multivariate}
Jasjeet~S Sekhon.
\newblock Multivariate and propensity score matching software with automated
  balance optimization: the matching package for r.
\newblock \emph{Journal of Statistical Software, Forthcoming}, 2008.

\bibitem[Sobel(2006)]{sobel_2006}
Michael~E Sobel.
\newblock What do randomized studies of housing mobility demonstrate?
\newblock \emph{Journal of the American Statistical Association}, 101\penalty0
  (476):\penalty0 1398--1407, 2006.

\bibitem[Stuart(2010)]{stuart2010matching}
Elizabeth~A Stuart.
\newblock Matching methods for causal inference: A review and a look forward.
\newblock \emph{Statistical science: a review journal of the Institute of
  Mathematical Statistics}, 25\penalty0 (1):\penalty0 1, 2010.

\bibitem[Sussman and Airoldi(2017)]{sussman2017elements}
Daniel~L Sussman and Edoardo~M Airoldi.
\newblock Elements of estimation theory for causal effects in the presence of
  network interference.
\newblock \emph{arXiv preprint arXiv:1702.03578}, 2017.

\bibitem[Sävje(2021)]{savje2021miss}
Fredrik Sävje.
\newblock Causal inference with misspecified exposure mappings, 2021.
\newblock URL \url{https://arxiv.org/abs/2103.06471}.

\bibitem[Tang et~al.(2010)Tang, Agarwal, O'Brien, and
  Meyer]{tang2010overlapping}
Diane Tang, Ashish Agarwal, Deirdre O'Brien, and Mike Meyer.
\newblock Overlapping experiment infrastructure: More, better, faster
  experimentation.
\newblock In \emph{Proceedings of the 16th ACM SIGKDD international conference
  on Knowledge discovery and data mining}, pages 17--26, 2010.

\bibitem[Tchetgen and VanderWeele(2012)]{tchetgen2012}
Eric J~Tchetgen Tchetgen and Tyler~J VanderWeele.
\newblock On causal inference in the presence of interference.
\newblock \emph{Statistical Methods in Medical Research}, 21\penalty0
  (1):\penalty0 55--75, 2012.

\bibitem[Toulis and Kao(2013)]{toulis2013estimation}
Panos Toulis and Edward Kao.
\newblock Estimation of causal peer influence effects.
\newblock In \emph{International conference on machine learning}, pages
  1489--1497. PMLR, 2013.

\bibitem[Traud et~al.(2012)Traud, Mucha, and Porter]{TRAUD20124165}
Amanda~L. Traud, Peter~J. Mucha, and Mason~A. Porter.
\newblock Social structure of facebook networks.
\newblock \emph{Physica A: Statistical Mechanics and its Applications},
  391\penalty0 (16):\penalty0 4165--4180, 2012.
\newblock ISSN 0378-4371.
\newblock \doi{https://doi.org/10.1016/j.physa.2011.12.021}.
\newblock URL
  \url{https://www.sciencedirect.com/science/article/pii/S0378437111009186}.

\bibitem[Ugander et~al.(2013)Ugander, Karrer, Backstrom, and
  Kleinberg]{ugander_graph_cluster}
Johan Ugander, Brian Karrer, Lars Backstrom, and Jon Kleinberg.
\newblock Graph cluster randomization: Network exposure to multiple universes.
\newblock In \emph{Proceedings of the 19th ACM SIGKDD International Conference
  on Knowledge Discovery and Data Mining}, KDD '13, page 329–337, New York,
  NY, USA, 2013. Association for Computing Machinery.
\newblock ISBN 9781450321747.

\bibitem[Viviano(2020)]{viviano2020experimental}
Davide Viviano.
\newblock Experimental design under network interference.
\newblock \emph{arXiv preprint arXiv:2003.08421}, 2020.

\bibitem[Vovk and Wang(2020)]{vovk2020combining}
Vladimir Vovk and Ruodu Wang.
\newblock Combining $p$-values via averaging.
\newblock \emph{Biometrika}, 107\penalty0 (4):\penalty0 791--808, 2020.

\bibitem[Wager and Xu(2021)]{wager2021experimenting}
Stefan Wager and Kuang Xu.
\newblock Experimenting in equilibrium.
\newblock \emph{Management Science}, 67\penalty0 (11):\penalty0 6694--6715,
  2021.

\bibitem[Xu et~al.(2015)Xu, Chen, Fernandez, Sinno, and
  Bhasin]{xu2015infrastructure}
Ya~Xu, Nanyu Chen, Addrian Fernandez, Omar Sinno, and Anmol Bhasin.
\newblock From infrastructure to culture: A/b testing challenges in large scale
  social networks.
\newblock In \emph{Proceedings of the 21th ACM SIGKDD International Conference
  on Knowledge Discovery and Data Mining}, pages 2227--2236, 2015.

\bibitem[Xu et~al.(2018)Xu, Duan, and Huang]{xu2018sqr}
Ya~Xu, Weitao Duan, and Shaochen Huang.
\newblock Sqr: Balancing speed, quality and risk in online experiments.
\newblock In \emph{Proceedings of the 24th ACM SIGKDD International Conference
  on Knowledge Discovery \& Data Mining}, pages 895--904, 2018.

\bibitem[Yates(1934)]{yates1934analysis}
Frank Yates.
\newblock The analysis of multiple classifications with unequal numbers in the
  different classes.
\newblock \emph{Journal of the American Statistical Association}, 29\penalty0
  (185):\penalty0 51--66, 1934.

\bibitem[Yu et~al.(2022)Yu, Airoldi, Borgs, and Chayes]{yu2022estimating}
Christina~Lee Yu, Edoardo~M Airoldi, Christian Borgs, and Jennifer~T Chayes.
\newblock Estimating the total treatment effect in randomized experiments with
  unknown network structure.
\newblock \emph{Proceedings of the National Academy of Sciences}, 119\penalty0
  (44):\penalty0 e2208975119, 2022.

\end{thebibliography}

\newpage

\appendix

\section{Simulation Details}
\label{section:simulation_detal}
\subsection{Under general assumptions}
In Section \ref{section:simulation_general}, we compare the power of the tests given in Algorithms~\ref{alg:test_one_exp}, \ref{alg:test_two_exp} and \ref{alg:test_three_exp}.

\subsubsection{Test statistics}
 Here, we discuss the test statistics used by the algorithms.
Let $H_{i,k}$ be the fraction of treated neighbors of unit $i$ in experiment $k$. Let $N_i$ be the number of neighbors of unit $i$ in the social network. 

\paragraph{One experiment.} For Algorithm~\ref{alg:test_one_exp}, we use the following test statistic: run a linear regression of 
\begin{equation}
\label{eqn:regression_stats_one}
Y_{\foc} \sim W_{\foc} + X_{\foc} + N_{\foc} + H_{\foc},
\end{equation}
extract the regression coefficient of $H$ and take the absolute value of the coefficient. 

\paragraph{Two experiments.} 
For Algorithm~\ref{alg:test_two_exp}, we consider two different test statistics, a correlation statistic and a regression statistic. For the correlation statistic, we take 
\begin{equation}
\label{eqn:corr_stats}
    T(W_{\foc,1:2}, X_{\foc}, Y^{\operatorname{diff}}_{\foc}, H_{\foc,1:2})
    = \abs{\Corr{Y^{\operatorname{diff}}_{\foc}, H_{\foc,2} - H_{\foc,1}}}.
\end{equation}
For the regression statistic, we run a regression of 
\begin{equation}
    Y^{\operatorname{diff}}_{\foc}
    \sim X_{\foc} + N_{\foc} + H_{\foc,1} + (H_{\foc,2} - H_{\foc,1}),
\end{equation}
extract the regression coefficient of $(H_{\foc,2} - H_{\foc,1})$ and take the absolute value of the coefficient. 

\paragraph{Three experiments.} 
Let $T_{k,l}$ be the test statistic (regression or correlation) defined above when only two experiments are utilized (the $k$-th and $l$-th experiments are utilized). We then simply use $T_{1,2} + T_{2,3} + T_{1,3}$ as the test statistic for Algorithm~\ref{alg:test_three_exp} with $K = 3$. 

\subsubsection{Outcome models}
We consider two different outcome models. For the linear model, let $H_{i,k}$ be the fraction of treated neighbors of unit $i$ in experiment $k$. We assume
\begin{equation}
\begin{split}
    Y_{i,k} &=  (\text{signal strength})H_{i,k} + 2W_{i,k} + X_{i,1} + X_{i,2} + \varepsilon_{i,k},
    \end{split}
\end{equation}
where $k \in \cb{1,2, 3}$ and $X_{i,1} \sim \mathcal{N}(0.5,1)$, $X_{i,2} \sim \text{Poisson}(3)$ independently. The errors $\varepsilon_{i,k}$'s are such that $(\varepsilon_{i,1}, \dots, \varepsilon_{i,K})$ is distributed as multivariate gaussian with $\EE{\varepsilon_{i,k}} = 0$, $\Var{\varepsilon_{i,k}} = 1$ and $\Cov{\varepsilon_{i,k}, \varepsilon_{i,l}} = (\text{fraction of common variance})$ for $k \neq l$. 

For the non-linear model, let $M_{i,k}$ be the number of treated neighbors of unit $i$ in experiment $k$. We assume 
\begin{equation}
\begin{split}
    Y_{i,k} &=  (\text{signal strength})\p{\frac{M_{i,k}}{20} + 5 \exp\p{\frac{1}{50}\min\p{M_{i,k}, 20}}} + \\
    & \qquad \qquad \qquad \qquad \qquad \qquad2W_{i,k} + X_{i,1} + X_{i,2} + \varepsilon_{i,k},
    \end{split}
\end{equation}
where $k \in \cb{1,2, 3}$ and $X_{i,1} \sim \mathcal{N}(0.5,1)$, $X_{i,2} \sim \text{Poisson}(3)$ independently. The errors $\varepsilon_{i,k}$'s are such that $(\varepsilon_{i,1}, \dots, \varepsilon_{i,K})$ is distributed as multivariate gaussian with $\EE{\varepsilon_{i,k}} = 0$, $\Var{\varepsilon_{i,k}} = 1$ and $\Cov{\varepsilon_{i,k}, \varepsilon_{i,l}} = (\text{fraction of common variance})$ for $k \neq l$.

\subsection{Time fixed effect model}
In Section \ref{section:simulation_time_fixed}, we compare the power of the tests given in Algorithms~\ref{alg:test_three_exp} and \ref{alg:hori_perm_multi}.

\subsubsection{Test statistics}
 Here, we discuss the test statistics used by the algorithms.
Let $H_{i,k}$ be the fraction of treated neighbors of unit $i$ in experiment $k$. Let $N_i$ be the number of neighbors of unit $i$ in the social network. 

\paragraph{Algorithm~\ref{alg:test_three_exp}.} We use the regression statistic defined in Section \ref{section:simulation_general}. 

\paragraph{Algorithm~\ref{alg:hori_perm_multi}.} 
For Algorithm~\ref{alg:hori_perm_multi}, we use an ``anova" statistic. Let $\mathcal{I}_1' = \cb{i\in \mathcal{I}_1: W_{i,1} = 1}$ and let $\mathcal{I}_m' = \cb{m(i): i\in \mathcal{I}_1'}$.
We start with concatenate $Y^{\operatorname{diff}}_{\operatorname{concat}} = \Big(Y^{\operatorname{diff}}_{\mathcal{I}_1',1}, Y^{\operatorname{diff}}_{\mathcal{I}_1,2}, Y^{\operatorname{diff}}_{\mathcal{I}_1,3}\Big)$. Similarly, let $N_{\operatorname{concat}} = (N_{\operatorname{concat},1}, N_{\operatorname{concat},m})$, where $N_{\operatorname{concat},1} = \Big(N_{\mathcal{I}_1',1}, N_{\mathcal{I}_1,2}, N_{\mathcal{I}_1,3}\Big)$ and $N_{\operatorname{concat},m} = \Big(N_{\mathcal{I}_1',1}, N_{\mathcal{I}_1,2}, N_{\mathcal{I}_1,3}\Big)$. 
We do the same concatenation for $X$ and $H$. The reason we take the subset $\mathcal{I}_1'$ of $\mathcal{I}_1$ in the first experiment is that we want $Y^{\operatorname{diff}}_{\operatorname{concat}}$ to be a pure contrast of treatment group and control group. Without the subsetting step,  $Y^{\operatorname{diff}}$ contains both treatment-control differences and control-control differences. 
Let $\operatorname{Ind}_2$ be the indicator of the second experiment and $\operatorname{Ind}_3$ be the indicator of the third experiment. We then run two regressions:
\begin{equation}
\begin{split}
&\text{Model 1: } Y^{\operatorname{diff}}_{\operatorname{concat}}
    \sim X_{\operatorname{concat}} + H_{\operatorname{concat}} + N_{\operatorname{concat}}
    + \operatorname{Ind}_2 + \operatorname{Ind}_3,\\
&\text{Model 2: } Y^{\operatorname{diff}}_{\operatorname{concat}}
\sim X_{\operatorname{concat}}  + N_{\operatorname{concat}}.
\end{split}
\end{equation}
Finally, we let the test statistic be the $F$-statistic from the anova testing of contrasting Model 1 with Model 2. 

\subsubsection{Matching algorithms}
\paragraph{Random matching.} We sample $m(i)$ uniformly at random without replacement. 

\paragraph{Covariate-based matching.} We use optimal matching based on the Mahalanobis distance of observed covariates and $N_i$ \citep{sekhon2008multivariate}.

\subsubsection{Outcome models}
We consider two different outcome models. For the linear model, let $H_{i,k}$ be the fraction of treated neighbors of unit $i$ in experiment $k$. We assume
\begin{equation}
\begin{split}
    Y_{i,k} &=  (\text{signal strength}) (2W_i+1) H_{i,k} + 2W_{i,k} + X_{i,1} + X_{i,2} + \varepsilon_{i,k},
    \end{split}
\end{equation}
where $k \in \cb{1,2, 3}$ and $X_{i,1} \sim \mathcal{N}(0.5,1)$, $X_{i,2} \sim \text{Poisson}(3)$ independently. The errors $\varepsilon_{i,k}$'s are such that $(\varepsilon_{i,1}, \dots, \varepsilon_{i,K})$ is distributed as multivariate gaussian with $\EE{\varepsilon_{i,k}} = 0$, $\Var{\varepsilon_{i,k}} = 1$ and $\Cov{\varepsilon_{i,k}, \varepsilon_{i,l}} = (\text{fraction of common variance})$ for $k \neq l$. 

For the non-linear model, let $M_{i,k}$ be the number of treated neighbors of unit $i$ in experiment $k$. We assume 
\begin{equation}
\begin{split}
    Y_{i,k} &=  (\text{signal strength})(2W_i+1) \p{\frac{M_{i,k}}{20} + 5 \exp\p{\frac{1}{50}\min\p{M_{i,k}, 20}}} \\
    &  \qquad  \qquad  + 2W_{i,k} + X_{i,1}X_{i,2} + \mathbbm{1}\cb{X_{i,1}>0.5, X_{i,2} > 3.5} + \varepsilon_{i,k},
    \end{split}
\end{equation}
where $k \in \cb{1,2, 3}$ and $X_{i,1} \sim \mathcal{N}(0.5,1)$, $X_{i,2} \sim \text{Poisson}(3)$ independently. The errors $\varepsilon_{i,k}$'s are such that $(\varepsilon_{i,1}, \dots, \varepsilon_{i,K})$ is distributed as multivariate gaussian with $\EE{\varepsilon_{i,k}} = 0$, $\Var{\varepsilon_{i,k}} = 1$ and $\Cov{\varepsilon_{i,k}, \varepsilon_{i,l}} = (\text{fraction of common variance})$ for $k \neq l$.

\end{document}